\documentclass{aa}
\usepackage{graphicx}
\usepackage{latexsym}
\usepackage{amsmath}
\usepackage{amssymb}

\newcommand{\aj}{AJ}
\newcommand\apj{{ApJ}}
\newcommand\aap{{A\&A}}
\newcommand\aaps{{A\&AS}}
\newcommand\pasp{{PASP}}

\newcommand{\MC}{\multicolumn}
\newcommand{\sunn}{$_{\odot}$}
\newcommand{\kms}{km~s$^{-1}$}

\begin{document}

   \title{SBS~0335--052~E and W:
	 implications of new broad-band and H$\alpha$ photometry}

   \author{S.A. Pustilnik\inst{1}\fnmsep\inst{3} \and
	   A.G. Pramskij\inst{1}\fnmsep\inst{3}  \and
	   A.Y. Kniazev\inst{1}\fnmsep\inst{2}\fnmsep\inst{3}
          }

   \offprints{S. Pustilnik, \email{sap@sao.ru}}

   \institute{
	     Special Astrophysical Observatory,
	     Nizhnij Arkhyz, Karachai-Circassia, 369167, Russia
	     \and
	     Max Planck Institut f\"{u}r Astronomie,
	     K\"{o}nigstuhl 17, D-69117, Heidelberg, Germany
	     \and
	     Isaac Newton Institute of Chile, SAO Branch
	     }

   \date{Received 22 July 2003; accepted 2 March 2004}

   \abstract{
We present the results of deep multicolour CCD imaging with the SAO RAS
6\,m telescope of the pair of extremely metal-deficient gas-rich dwarf
galaxies SBS~0335--052~E and W.
The total magnitudes in $U,B,V,R$ and $I$ bands and the integrated fluxes
of H$\alpha$ emission are measured for both galaxies, and their integrated
colours are derived.
The analysis of their surface brightness (SB) distributions is performed
with the use of the azimuthally-averaged SB profiles. The latter were modeled
by the central Gaussian component and the underlying exponential `disk',
mainly contributing in the outermost, very low SB regions. The colours of
these LSB components are used to
estimate the age of the oldest visible  stellar population.
For the interpretation of the observed LSB colours their contamination by
the nebular emission of ionized gas is accounted for by the use of the
distribution of H$\alpha$ flux.
We compare the derived `gas-free' colours with the colours predicted by the
evolution synthesis models from the PEGASE.2 package, considering three SF
histories: a) instantaneous  starburst, b) continuous star formation with
constant SFR, and c) continuous exponentially fading SF (e-fold time of 3
Gyr). We conclude that the `gas-free' colours of the LSB component of the
Eastern galaxy can be best consistent with the instantaneous starburst
population (ages of $\lesssim$100 Myr). Models with continuous SF give less
consistent results, but can be considered as acceptable.
For such scenarios, the `gas-free' colours
require ages of $\lesssim$400 Myr. For the Western galaxy, the situation
is in general similar. But the colour $(V-R)$ appears to be quite red and
implies a significantly older component. We briefly discuss the possible
evolution sequence between SBS 0335--052, HI 1225+01 and other extremely
metal-deficient galaxies based on the merger scenario.
	    }

   \authorrunning{S.Pustilnik et al.}

   \titlerunning{Broad-band and H$\alpha$ photometry of SBS~0335--052~E and W}

   \maketitle

\section{Introduction}

The system SBS~0335--052 E,W -- a unique pair of dwarf galaxies with the
projected
separation of 22 kpc, and the lowest (together with I~Zw~18) ionized gas
oxygen abundances known (1/42 and 1/50 of solar value,
respectively)\footnote{The solar value is accepted here
as 12+$\log$(O/H)=8.92 (Anders \& Grevesse \cite{AG89}) to discus
the oxygen abundance in the same system as other authors. In fact,
recent data show that this value seems to be 0.2 dex lower (Prieto et al.
\cite{Prieto01}).} was first described by Izotov et al. (\cite{Izotov90}),
Pustilnik et al. (\cite{Pustilnik97}) and Lipovetsky et al.
(\cite{Lipovetsky99}).
This galaxy pair has been intensively studied over the last 13 years, and
there are about 40 papers devoted to various aspects of this multiwavelength
and multimethod attack. Here we concentrate mainly on its photometric
properties, related to its age, postponing the discussion of some other
properties of this system to Sect. \ref{Discussion}.

The very blue
$(V-I)$ colour of SBS 0335--052~E was first derived from the HST photometry by
Thuan et al. (\cite{TIL97}). Also, in this work the starburst was resolved
into several superstar clusters.
In papers by Izotov et al. (\cite{ILC97,Izotov01}) it was shown that the
H$\alpha$ emission of this galaxy is very extended.
It was argued that this nebular emission is mixed with and well diluted by
the radiation of the sea of unresolved underlying late B and early A stars.

The detailed analysis of the broad-band surface photometry of this galaxy
coupled with the high S/N long-slit spectra was presented by Papaderos et al.
(\cite{Papa98}). The broad-band colours of the LSB halo were modeled by
the mixture of stellar and gas emission. The observed nebular emission was
directly accounted for in this analysis for the region covered by the
long-slit strip of
1\arcsec$\times$6\arcsec. These authors concluded that their data indicate
no detectable contribution of old stars to the LSB underlying halo. However,
Kunth \& \"Ostlin (\cite{KO_01}) questioned their results, mainly due to the
use of the different models for colour evolution.

The extremely low oxygen abundance of the Western galaxy was determined by
Lipovetsky et al. (\cite{Lipovetsky99}). Its first $R,I$ photometry was also
presented in this paper.
Papaderos et al. (\cite{Papa98}) presented the detailed analysis of the
total and SB broad-band photometry of this galaxy. While their photometry was
rather deep, the compactness and faintness of this galaxy
prevented to perform the analysis of its evolution status.
Near IR data on SBS 0335--052~E and W colours were presented by Vanzi et al.
(\cite{Vanzi00}). Despite that observations in the NIR range are the most
sensitive to the presence of an old stellar population, only the upper limit
($\sim$15\%) on the contribution of its NIR radiation in the Eastern galaxy
was derived.

Up to now, the absence of very deep optical photometry supported by a deep
imaging in nebular emission lines (first of all, in H$\alpha$), or even better
with full two-dimensional spectrophotometry, prevented us from getting
reliable
colours of an underlying ``old'' stellar population in this type of object.
The essence of this new step, performed with the SAO 6\,m telescope data, is
an attempt to elaborate methods for similar
studies of other actively studied or recently discovered
extremely metal-deficient (hereafter XMD) BCGs (e.g., Papaderos
et al. \cite{Papa99}; Kniazev et al. \cite{Kniazev00a,Kniazev00b}; Fricke et
al. \cite{Fricke01}; Guseva et al. \cite{Guseva01};  Melbourne \& Salzer
\cite{MS02};
Ugryumov et al. \cite{HSS_LM}; Pustilnik et al. \cite{HS0837}).
A similar approach was recently realized for the analysis of the LSB
underlying component in I~Zw~18 by Papaderos et al. (\cite{Papa02}).
In Sect. \ref{Obs} we describe the observations and their reduction.
Sect. \ref{Results} presents the results of the data analysis.
We further compare the derived colours with the model tracks of evolving
stellar populations in Sect. \ref{models}.
These results are discussed in Sect. \ref{Discussion}.
We summarize our findings and draw conclusions in Sect. \ref{conc}.
All distance-dependent parameters discussed in the paper are
calculated with the adopted distance to the system of 54.3 Mpc.
This corresponds to the scale of 263 pc in 1\arcsec.

\section{Observations and data reduction}
\label{Obs}

$UBVRI$ broad-band and H$\alpha$ narrow-band photometry was
carried out with the 6\,m telescope (BTA) of the Special Astrophysical
Observatory of Russian Academy of Sciences (SAO RAS) during
three runs in November--December 1997 (Table~\ref{tabObs}).
Observational data were obtained in the prime focus of the telescope using
the
1040$\times$1160 pixel ISD017A CCD detector, binned 2$\times$2,
 with the gain $2.3$ e/ADU and the resulting scale 0.274
arcsec~pixel$^{-1}$. The broad-band $U$, $B$ and $V$ are the Johnson
photometry
system filters, while $R$ and $I$  filters are those from the Cousins system
(Bessell \cite{Bessell90}).
Observations in H$\alpha$ were performed in two narrow-band filters with
FWHM = 85~\AA,  centered at $\lambda$6660~\AA\ (H$\alpha$-line) and
$\lambda$6740~\AA\ (H$\alpha$-cont) for measurements of the H$\alpha$
emission line and H$\alpha$-free continuum flux, respectively.
Observations were
conducted with the software package {\tt NICE} in
MIDAS\footnote{MIDAS is an acronym for the European Southern
Observatory package --- Munich Image Data Analysis System.},
as described by Kniazev \& Shergin (\cite{Kniazev95}).
The exposure times, mean values of airmass and  seeings for
observations in each filter are presented in Table~\ref{tabObs}.
The images in each filter were
obtained by summing several exposures. The individual exposures
in broad-band filters ($BVI$), as well as those for H$\alpha$ narrow filters,
were conducted in a cyclic sequence
($B\Rightarrow$$V\Rightarrow$$I\Rightarrow$$B$). Each night we obtained
bias, dark and flat-field images for primary reduction.
The broad-band photometric standards were observed in the fields PG~0039+049,
Feige~22, G93--48 and PG~2317+046  (Landolt \cite{Landolt92}).
For calibration of narrow-band H$\alpha$ frames, we observed spectrophotometric
standards BD+28\degr4211, Feige~11 and Feige~34 from Bohlin
(\cite{Bohlin96}).

Standard reduction steps, including bias-subtraction, flat-field
correction, removal of cosmic ray hits and absolute flux calibration
were carried out using
IRAF\footnote{IRAF: the Image Reduction and Analysis Facility is
distributed by the National Optical Astronomy Observatories, which is
operated by the Association of Universities for Research in Astronomy,
In. (AURA) under cooperative agreement with the National Science
Foundation (NSF).}
and MIDAS.
The aperture photometry of SBS~0335--052~E and W and of the standard stars
was performed in IRAF with {\tt APPHOT} tasks {\tt polyphot} and {\tt phot}.
The surface brightness profiles were obtained with the {\tt STSDAS} task
{\tt ellipse}.

\begin{table}
\label{tabObs}
\centering
\caption{Journal of Observations}
\begin{tabular}{clccc} \\ \hline \hline
\MC{1}{c}{Date} & \MC{1}{c}{Band} & \MC{1}{c}{Exposure} & \MC{1}{c}{Seeing}   & \MC{1}{c}{Airmass} \\
		&                 & \MC{1}{c}{time [s]} & \MC{1}{c}{[arcsec]} &                    \\
\MC{1}{c}{(1)}  & \MC{1}{c}{(2)}  & \MC{1}{c}{(3)}      & \MC{1}{c}{(4)}      & \MC{1}{c}{(5)}     \\
\hline\hline
\\[-0.3cm]
09.11.1997      & H$\alpha$-line  & 6$\times$600        &  1.9                &       1.53         \\
09.11.1997      & H$\alpha$-cont  & 6$\times$600        &  1.9                &       1.53         \\
09.11.1997      & $R_\mathrm{c}$  & 3$\times$120        &  1.9                &       1.55         \\
26.11.1997      & $B$             & 3$\times$300        &  2.3                &       1.54         \\
26.11.1997      & $V$             & 3$\times$300        &  2.2                &       1.54         \\
26.11.1997      & $R_\mathrm{c}$  & 1$\times$300        &  2.0                &       1.54         \\
26.11.1997      & $I_\mathrm{c}$  & 3$\times$300        &  1.9                &       1.54         \\
01.12.1997      & $U$             & 4$\times$900        &  3.0                &       1.55         \\
\\[-0.3cm]
\hline \hline \\[-0.2cm]
\end{tabular}
\end{table}

\section{Results}
\label{Results}

   \begin{figure*}
   \centering
   \includegraphics[width=7.4cm]{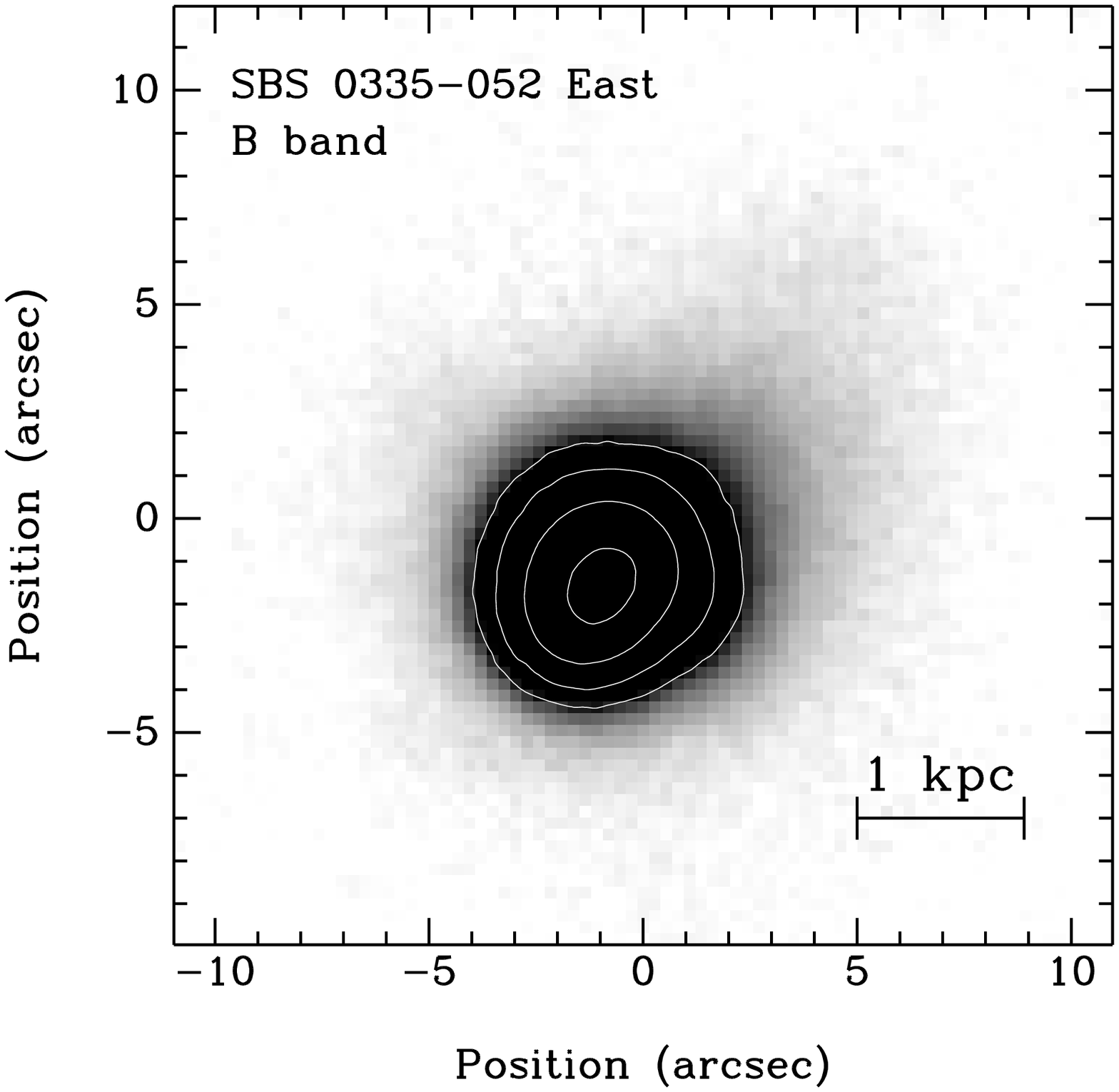}
   \includegraphics[width=7.4cm]{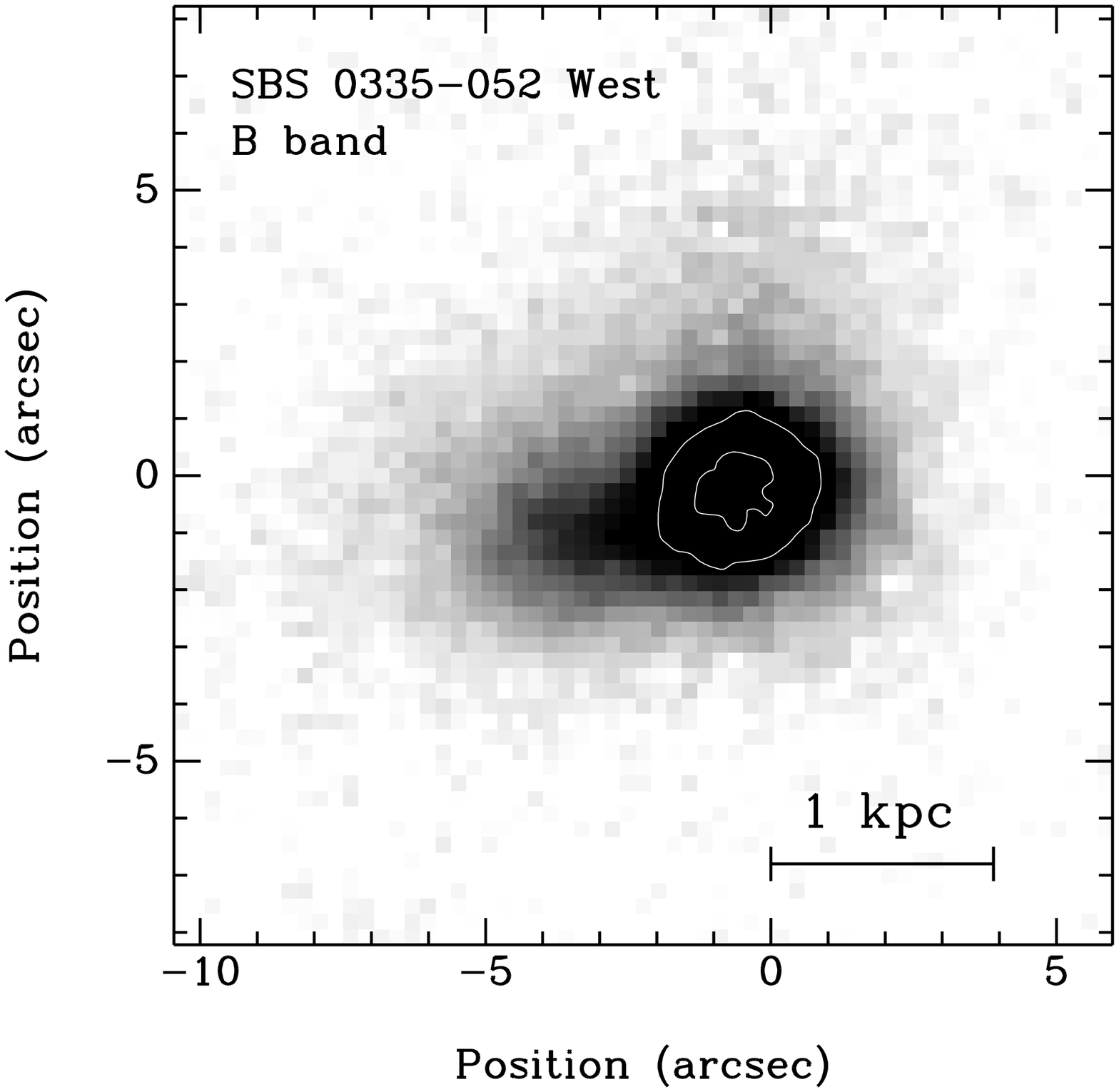}
      \caption{BTA B-images of SBS~0335--052 East (left) and West (right)
      galaxies in grey-scale. The superimposed isophotes from the center
      outwards correspond to surface brightness levels of 19.7, 20.7, 21.7
      and 22.5 mag~arcsec$^{-2}$ - for the east galaxy, and 22.3 and 22.9
      mag~arcsec$^{-2}$ - for the west galaxy.
              }
	 \label{Image_B}
   \end{figure*}

\subsection{Integrated photometry}

\begin{table}
\caption[]{Total magnitudes and colours}
\label{tabTotMag}
\begin{tabular}{lrrc} \hline\hline
Parameter               & 0335$-$052~E    & 0335$-$052~W   & Ref \\
\hline
\\[-0.3cm]
$U$$^{\ddagger}$        & $16.24\pm0.06$  & $18.71\pm0.06$ & 1    \\
			& $16.24\pm0.04$  & $18.74\pm0.08$ & 2    \\
$B$                     & $16.95\pm0.03$  & $19.14\pm0.03$ & 1    \\
			& $16.96\pm0.02$  & $19.29\pm0.07$ & 2    \\
$V$                     & $16.65\pm0.03$  & $19.25\pm0.03$ & 1    \\
			& $16.65\pm0.01$  &   ........~~~~ & 3    \\
$R$                     & $16.52\pm0.03$  & $18.82\pm0.04$ & 1    \\
			& $16.57\pm0.01$  & $19.03\pm0.05$ & 4    \\
$I$                     & $16.88\pm0.04$  & $18.94\pm0.04$ & 1    \\
			& $16.88\pm0.01$  & $19.08\pm0.05$ & 3,4  \\
\\[-0.3cm]
$(U-B)$                 & $-0.71\pm0.07$  & $-0.43\pm0.07$ & 1    \\
			& $-0.72\pm0.05$  & $-0.54\pm0.10$ & 2    \\
$(B-V)$                 & $0.30\pm0.05 $  & $-0.12\pm0.05$ & 1    \\
			& $0.31\pm0.03 $  &   ........~~~~ & 2,3     \\
$(V-R)$                 & $0.13\pm0.05 $  & $0.44\pm0.05 $ & 1    \\
			& $0.08\pm0.02 $  &   ........~~~~ & 3,4     \\
$(V-I)$                 & $-0.23\pm0.05$  & $0.31\pm0.05 $ & 1    \\
			& $-0.23\pm0.02$  &   ........~~~~ & 3     \\
$(R-I)$                 & $-0.36\pm0.05$  & $-0.12\pm0.05$ & 1    \\
			& $-0.35\pm0.02$  & $-0.05\pm0.07$ & 2,4     \\
\\[-0.3cm]
F(H$\alpha$)$^{\dag}$   & $3227\pm107  $  & $144\pm6$~~~~~ & 1    \\
\\[-0.3cm]
\hline\hline
\\[-0.3cm]
\multicolumn{4}{l}{$^{\ddagger}$ Adopted for the east galaxy equal to that in (2);} \\
\multicolumn{4}{l}{$^{\dag}$ Total flux in H$\alpha$-line in units 10$^{-16}$ erg~cm$^{-2}$s$^{-1}$;}  \\
\multicolumn{4}{l}{(1) -- this paper; (2) -- Papaderos et al. (\cite{Papa98});} \\
\multicolumn{4}{l}{(3) -- Thuan et al. (\cite{TIL97}); (4) -- Lipovetsky et al. (\cite{Lipovetsky99}); } \\
\end{tabular}
\end{table}

In Table \ref{tabTotMag} we present the total magnitudes and colours for the
east and west galaxies of the SBS~0335--052 system. For comparison, we show in
this
table the same parameters derived by Thuan et al. (\cite{TIL97}),
Papaderos et al. (\cite{Papa98}) and Lipovetsky et al.
(\cite{Lipovetsky99}). The errors of our instrumental magnitudes are very
small for both east and west galaxies, less than $\sim$0\fm01, and are
significantly lower than the uncertainties resulting from the photometric
system transformations. Therefore, despite the large difference in the total
magnitudes of the east and west galaxies, the errors of their total
magnitudes are very similar.
For the $U$-band we had no photometric system parameters. We observed two
standard stars, one of which was at different air mass. From these
observations
we estimated the extinction and the colour terms, and got the first
approximation for the total $U$-band magnitude of the east galaxy,
0\fm14 brighter
than that from Papaderos et al. (\cite{Papa98}). Since the accuracy of
the calibration of our $U$-band data was difficult to assign, we adopted
the total $U$-band magnitude for the east galaxy as equal
to that from Papaderos et al. (\cite{Papa98}).
For the west galaxy the $U$-magnitude
was simply recalculated from the measured instrumental magnitude difference
between the east and west galaxies.
The difference $\Delta U$ for the west galaxy between our and the
Papaderos et al.
values is significantly lower than 1$\sigma$ of the combined error.

In $B$-band our integrated magnitude of the east galaxy coincides nicely
with the value of Papaderos et al (\cite{Papa98}).
For the west galaxy the difference is larger,
but still acceptable, approximately within 2$\sigma$ of the combined error.
In the $V$-band there exist in the literature only integrated magnitudes for
the east galaxy (Thuan et al. \cite{TIL97}). It exactly coincides with our
measurement. The total magnitudes in $R$-band were published by Lipovetsky et
al. (\cite{Lipovetsky99}). Our value for the east galaxy is 0\fm05 brighter
than
theirs, but still is well consistent (within 1.7$\sigma$ of the combined
error).
However, for the west component we measure the total magnitude to be
0\fm21 brighter than the value in Lipovetsky et al. (\cite{Lipovetsky99}).
This difference exceeds 3$\sigma$ of the
combined error ($\sigma_{\rm comb}$=0\fm064), and indicates possible
systematic effects. A similar situation, but less striking, occurs for
the $I$-band. Again, for the east galaxy our total magnitude and that from
Thuan et al. (\cite{TIL97}) coincide completely. However, for the west galaxy,
our total $I$-band magnitude is 0\fm14 brighter than that from Lipovetsky
et al. (\cite{Lipovetsky99}). The combined error for this difference is the
same as for the $R$-band,
and this difference slightly exceeds 2$\sigma_{\rm comb}$.

Thus, besides $R$ and $I$ total magnitudes for the west galaxy, our data are
consistent with previous measurements. For the latter, the difference
comes from the results obtained with the Calar Alto 2.2-m telescope, with the
limited photon statistics for the west galaxy (compare their reduced S/N for
the west galaxy relative to that of the east galaxy). This implies that small
uncertainties
in the determination of the sky background could additionally affect the
integrated magnitude of the west galaxy, and thus cause the slight decrease in
the flux in $R$ and $I$. Probably, a similar effect in the $B$-band can
explain
the difference (0\fm15)  of our and Papaderos et al. magnitudes for the west
galaxy.

The derivation of H$\alpha$-flux could be affected by the emission lines that
appeared within the H$\alpha$-cont filter passband and the small variations
of the
CCD detector sensitivity between the wavelengths of the two narrow filters.
The former is mostly due to H$\alpha$ itself in the H$\alpha$-cont filter  and
the \ion{He}{i}-line $\lambda$6678~\AA, which result in the decrease of the
measured H$\alpha$-flux by $\sim$3.8\% and $\sim$0.9\%, respectively. The
[\ion{S}{ii}]-lines
$\lambda\lambda$6717,6731~\AA\ contribute $\sim$0.3\%. Thus, the total
correction of 5\% to the directly measured H$\alpha$-flux was applied. The
small variations of the CCD detector sensitivity between the two narrow
filters are completely accounted for by observations of spectrophotometric
standards in each of them.
The measurements of H$\alpha$ flux for the east galaxy were presented by
Melnick
et al. (\cite{Melnick92}) (integrated over the region of 3 arcsec$^2$) and
for the west galaxy by Lipovetsky et al. (\cite{Lipovetsky99}) (integrated over
the region of 4 arcsec$^2$).
Our integrated H$\alpha$ flux for the east galaxy is a factor of $\sim$1.6
higher than that obtained by Melnick et al. (\cite{Melnick92}).
For the west galaxy our integrated H$\alpha$ flux is a factor of 2 higher than
that
obtained by Lipovetsky et al. (\cite{Lipovetsky99}). This implies that besides
the emission from the two compact knots, there is a significant contribution
from  more diffuse gas. The implications of the H$\alpha$ luminosity data
are discussed in Sect. \ref{SFR}.

\subsection{Surface photometry}
\label{surface}

\begin{table*}
\centering
\caption[]{Model parameters of the surface brightness profiles for SBS 0335--052 E and W}
\label{tabStrParam}
\begin{tabular}{lcccccccc} \hline \hline
Band & $\mu_\mathrm{0}$ & $\alpha_\mathrm{0}$  & $\alpha_\mathrm{0}$ & $\mu_\mathrm{G}$   & $\alpha_\mathrm{G}$ &$m_{\rm disk}$& $m_{\rm burst}$  & Ref \\
     & mag\,arcsec$^{-2}$&  arcsec   &  pc      & mag\,arcsec$^{-2}$&    arcsec  & mag & mag &     \\
 1   &         2         &      3    &  4      &      5             &     6      &  7  &  8  &  9   \\
\hline
\multicolumn{9}{c}{\bf {Eastern galaxy}} \\
$U$  & 20.27$\pm$0.05   &1.82$\pm$0.02 & 468$\pm$5~ & 20.10$\pm$0.02   & 3.88$\pm$0.03 &16.98$\pm$0.05 &17.01 & 1    \\
     & 20.66$\pm$0.08   &1.74$\pm$0.04 & 459$\pm$12 & ---              & ---           & ---           & ---  & 2    \\
$B$  & 21.27$\pm$0.04   &1.83$\pm$0.02 & 470$\pm$5~ & 19.90$\pm$0.02   & 2.83$\pm$0.03 &17.96$\pm$0.05 &17.49 & 1    \\
     & 21.38$\pm$0.05   &1.74$\pm$0.03 & 458$\pm$8~ & ---              & ---           & ---           & ---  & 2    \\
$V$  & 20.72$\pm$0.05   &1.72$\pm$0.02 & 442$\pm$5~ & 19.46$\pm$0.03   & 2.56$\pm$0.03 &17.55$\pm$0.05 &17.28 & 1    \\
     & 20.71$\pm$0.26   &1.43          & 376$\pm$30 & ---              & ---           & ---           & ---  & 3    \\
$R$  & 20.54$\pm$0.03   &1.71$\pm$0.02 & 439$\pm$5~ & 19.25$\pm$0.02   & 2.43$\pm$0.02 &17.38$\pm$0.04 &17.18 & 1    \\
     & 21.61$\pm$0.10   &2.32          & 610$\pm$15 & ---              & ---           & ---           & ---  & 2,4    \\
$I$  & 21.24$\pm$0.08   &1.89$\pm$0.04 & 486$\pm$10 & 19.43$\pm$0.03   & 2.38$\pm$0.03 &17.86$\pm$0.09 &17.44 & 1    \\
     & 20.96$\pm$0.35   &1.50          & 396$\pm$43 &  ---             & ---           & ---           & ---  & 2,3    \\
H$\alpha$ & 21.27$\pm$0.02 &1.61$\pm$0.01 &391$\pm$3~ & 20.02$\pm$0.02 & 2.16$\pm$0.02 &18.24$\pm$0.03 &18.15  & 1    \\
\hline
\multicolumn{9}{c}{\bf {Western galaxy} } \\
$U$  & 21.85$\pm$0.08   &1.62$\pm$0.04 & 416$\pm$10 & ---              & ---           &18.81$\pm$0.09 &21.35 & 1    \\
     & 22.27$\pm$0.34   &1.53$\pm$0.15 & 402$\pm$40 & ---              & ---           & ---           & ---  & 2    \\
$B$  & 22.49$\pm$0.10   &1.73$\pm$0.05 & 445$\pm$13 & 23.61$\pm$0.15   & 2.72$\pm$0.18 &19.30$\pm$0.12 &21.27 & 1    \\
     & 22.33$\pm$0.22   &1.47$\pm$0.07 & 387$\pm$20 & ---              & ---           & ---           & ---  & 2    \\
$V$  & 22.41$\pm$0.05   &1.61$\pm$0.03 & 414$\pm$5~ & 23.71$\pm$0.14   & 2.12$\pm$0.11 &19.38$\pm$0.06 &21.61 & 1    \\
$R$  & 22.06$\pm$0.03   &1.69$\pm$0.02 & 434$\pm$13 & 22.98$\pm$0.07   & 2.04$\pm$0.13 &18.93$\pm$0.04 &21.40 & 1    \\
$I$  & 22.20$\pm$0.08   &1.65$\pm$0.05 & 424$\pm$13 & 22.94$\pm$0.11   & 1.94$\pm$0.11 &19.12$\pm$0.09 &20.99 & 1    \\
H$\alpha$ & 22.82$\pm$0.10 &0.96$\pm$0.02 & 247$\pm$5 & ---            & ---           &20.91$\pm$0.12 &23.70 & 1    \\
\hline\hline
\\[-0.2cm]
\multicolumn{9}{l}{All values are not corrected for the Galactic extinction; Ref: (1) -- Data from this paper; }\\
\multicolumn{9}{l}{(2) -- Papaderos et al. (\cite{Papa98});  (3) -- Thuan et al. (\cite{TIL97}); (4) -- Lipovetsky et al. (\cite{Lipovetsky99});} \\
\end{tabular}
\end{table*}

   \begin{figure*}
   \centering
   \includegraphics[width=5.9cm,bb=33 248 583 709,clip=]{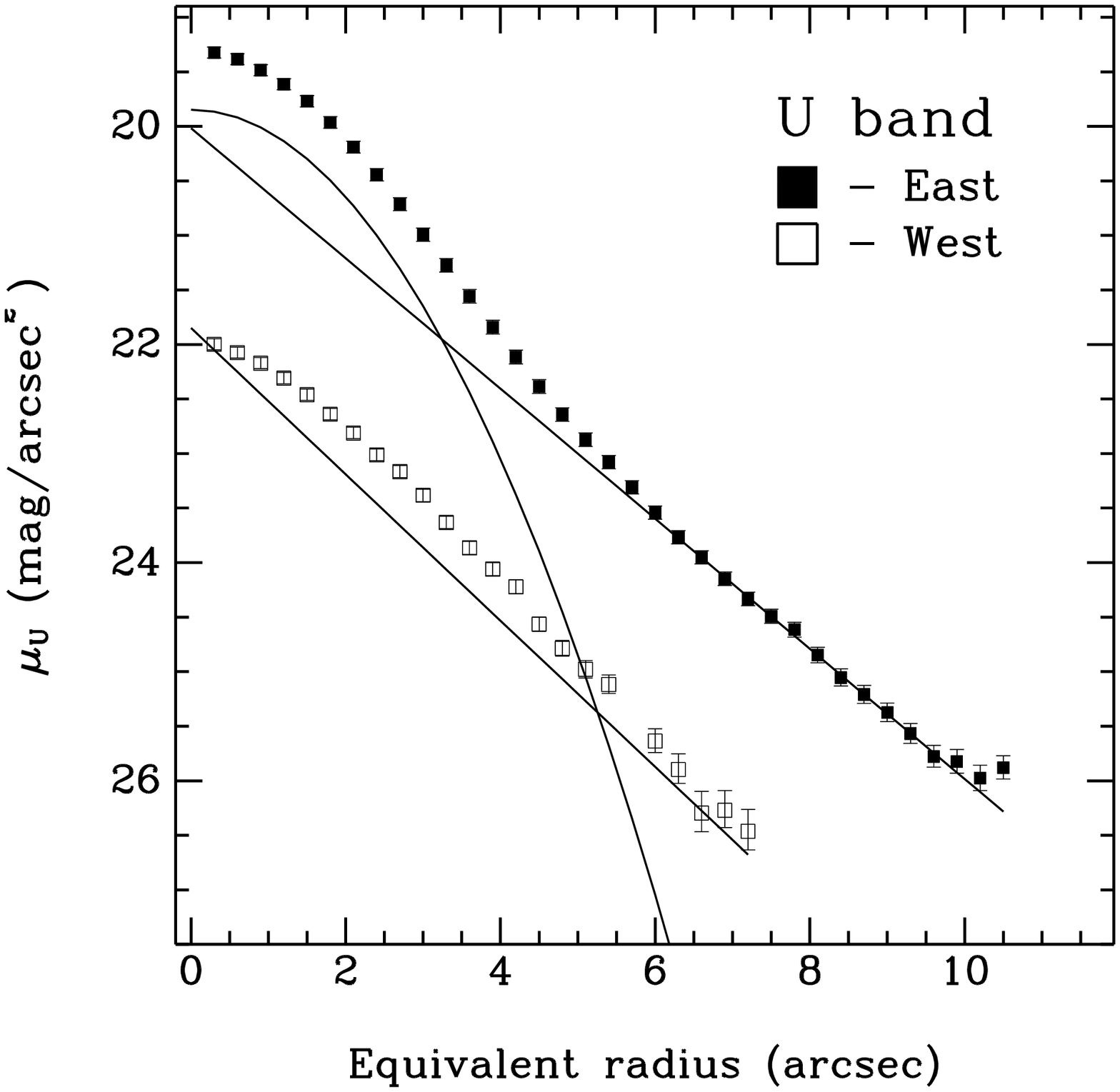}
   \includegraphics[width=5.9cm,bb=33 248 583 709,clip=]{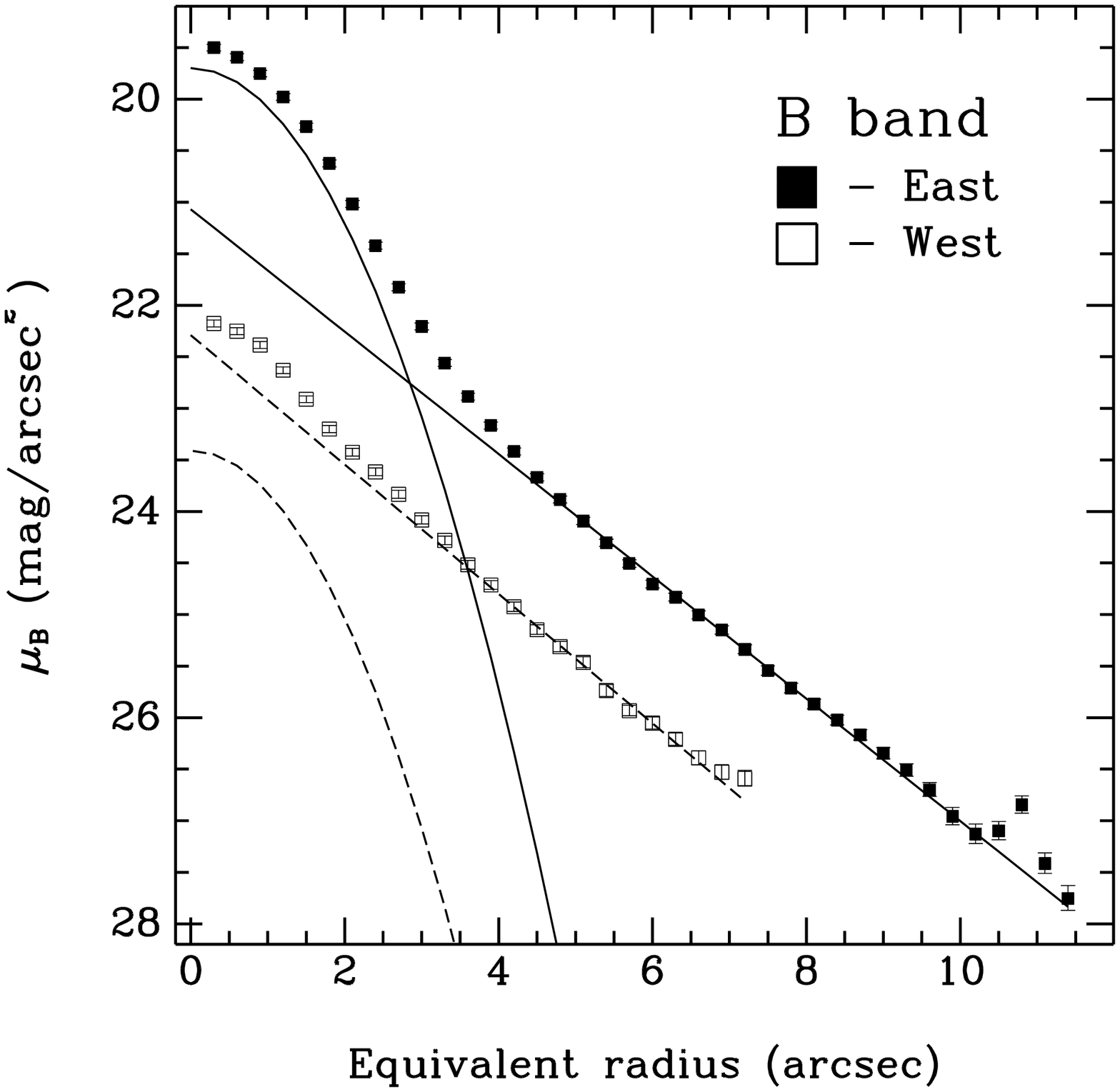}
   \includegraphics[width=5.9cm,bb=33 248 583 709,clip=]{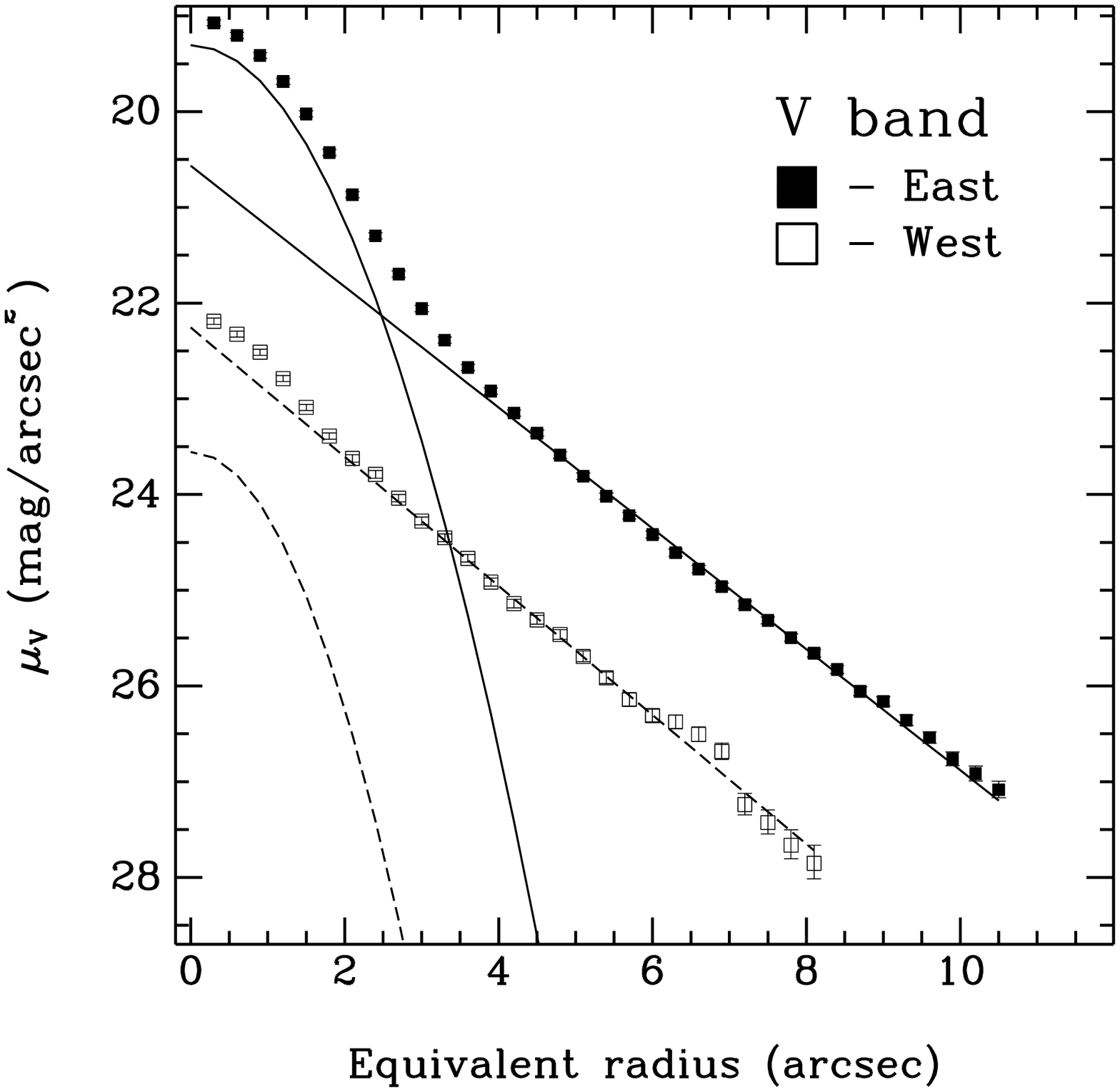}
   \includegraphics[width=5.9cm,bb=33 164 583 709,clip=]{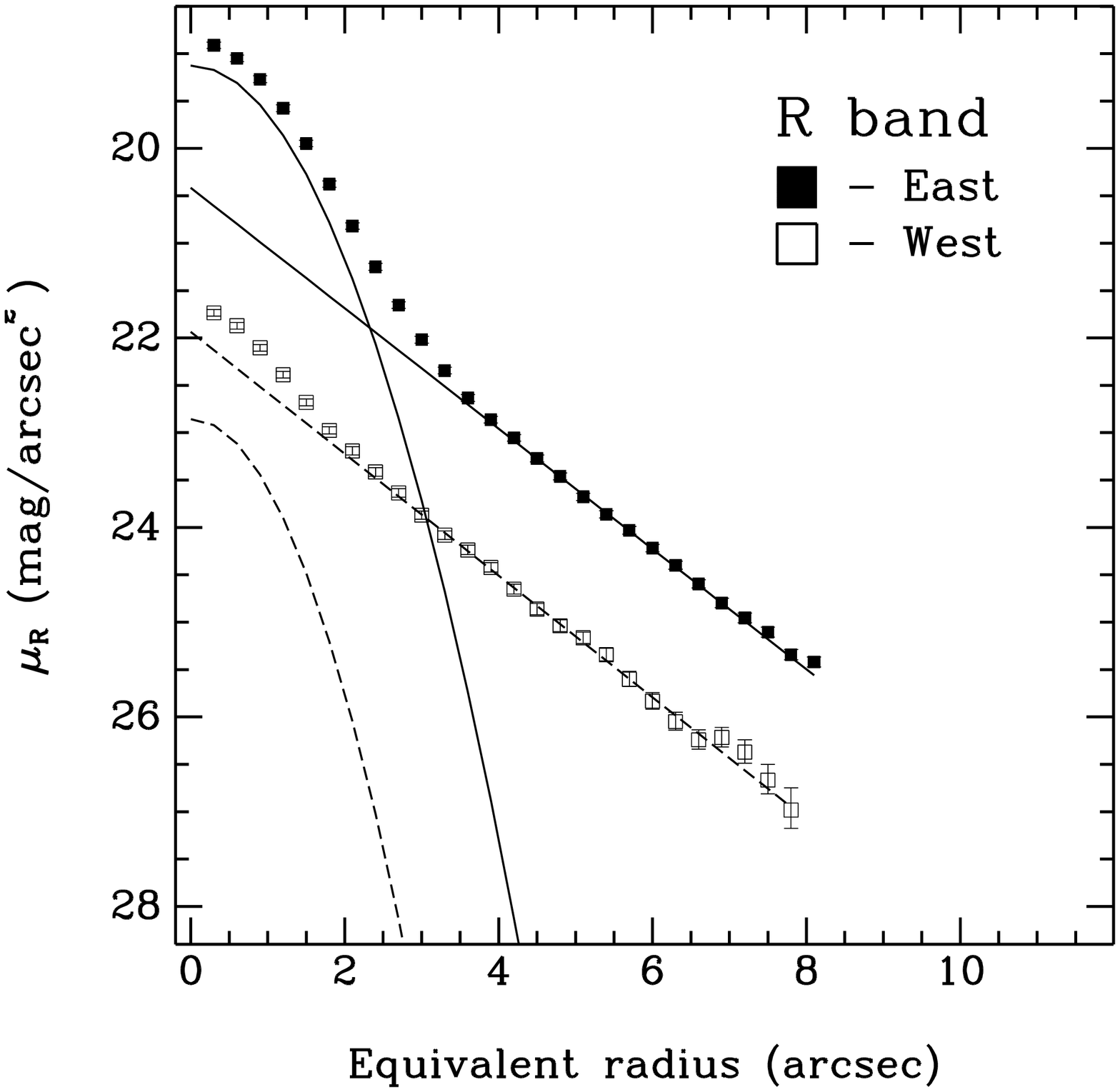}
   \includegraphics[width=5.9cm,bb=33 164 583 709,clip=]{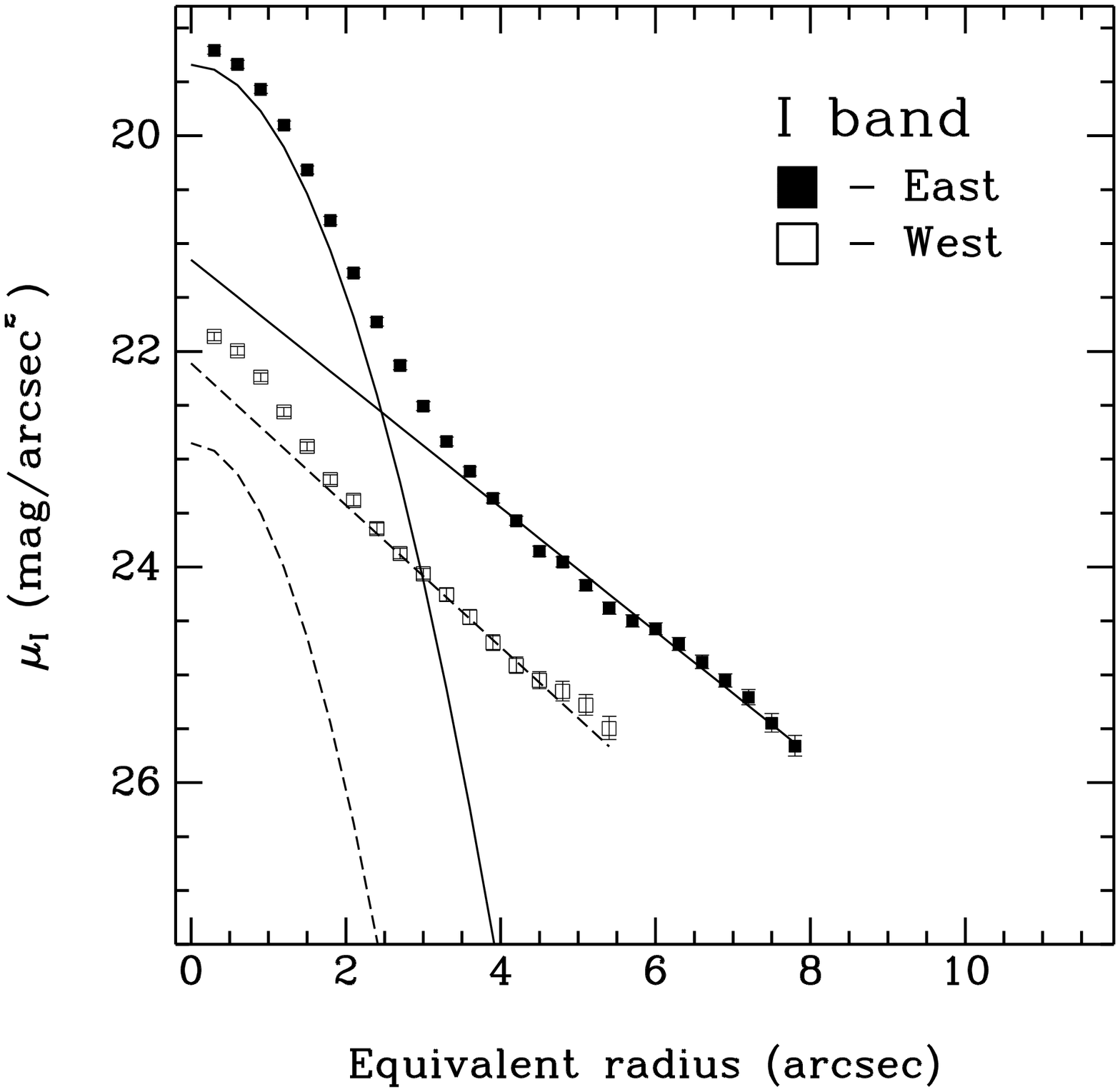}
   \includegraphics[width=5.9cm,bb=33 164 583 709,clip=]{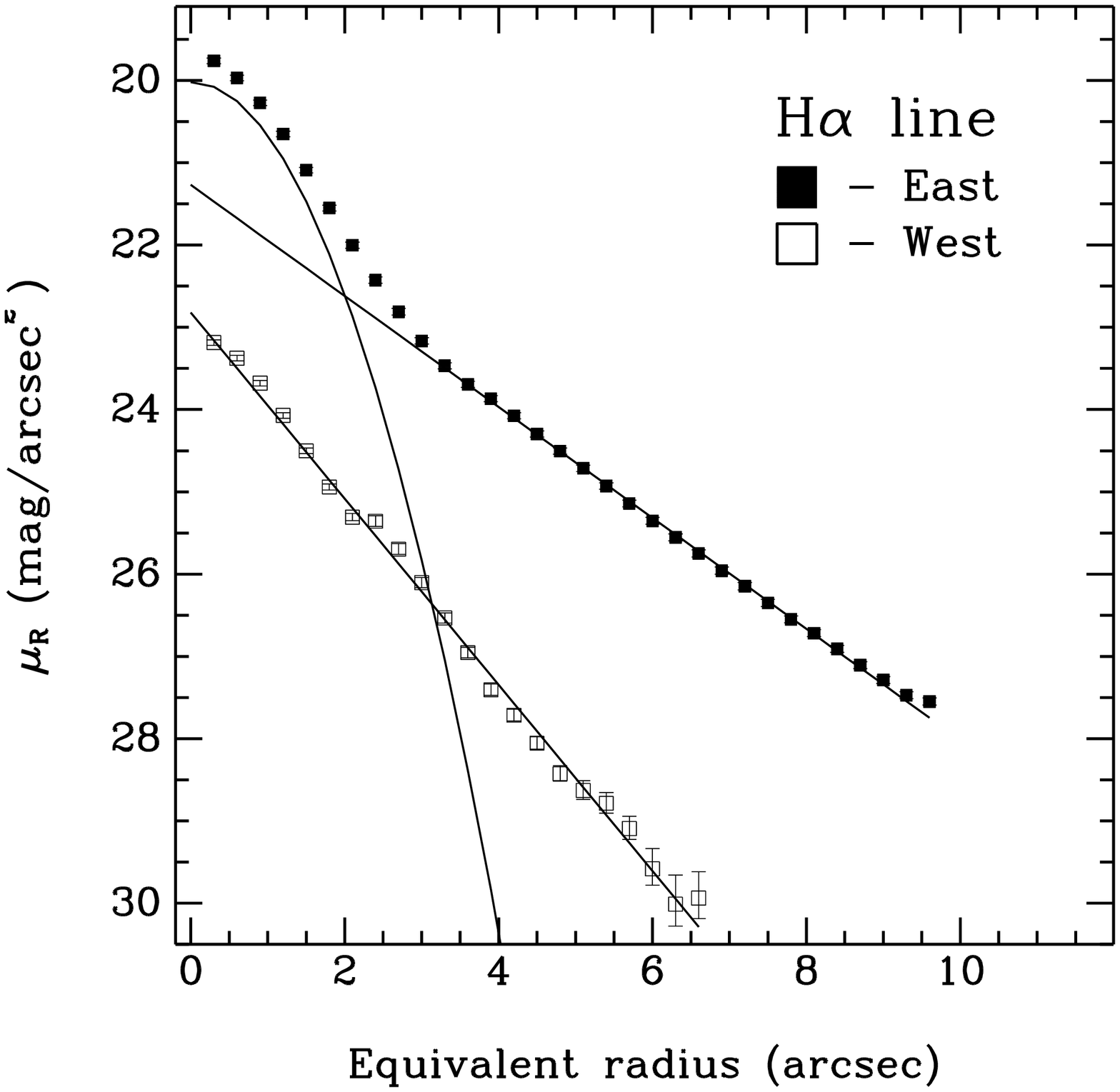}
      \caption{Surface brightness profiles (SBPs) for SBS~0335--052~E
	       and W. Instrumental H$\alpha$-line SBP
	       was transformed into conditional magnitudes using the $R$-band
	       photometrical coefficients. All the data
	       are corrected for the Galactic extinction.
              }
	 \label{FigSBP}
   \end{figure*}

   \begin{figure*}
   \centering
   \includegraphics[width=5.9cm,bb=33 248 583 709,clip=]{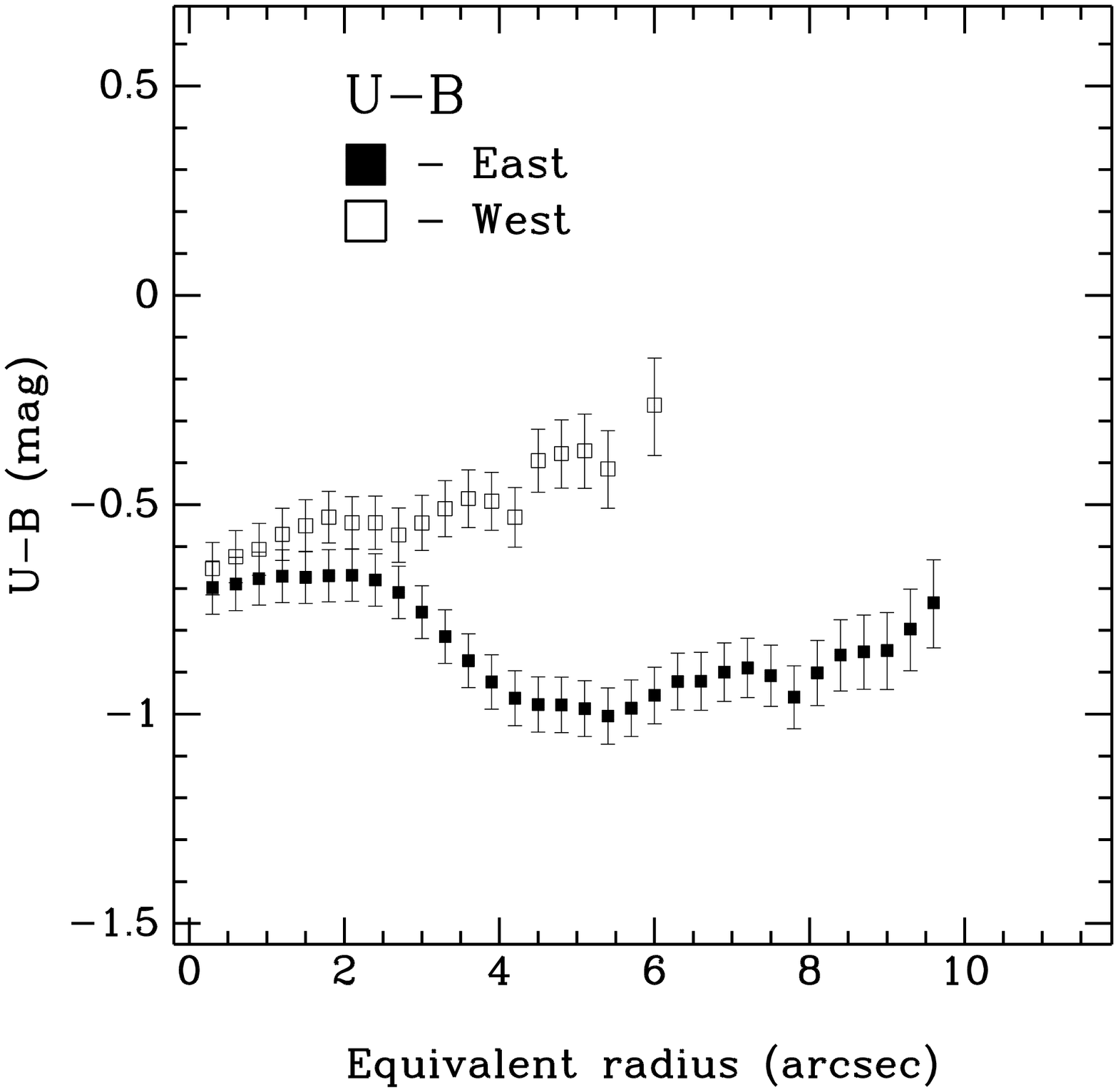}
   \includegraphics[width=5.9cm,bb=33 248 583 709,clip=]{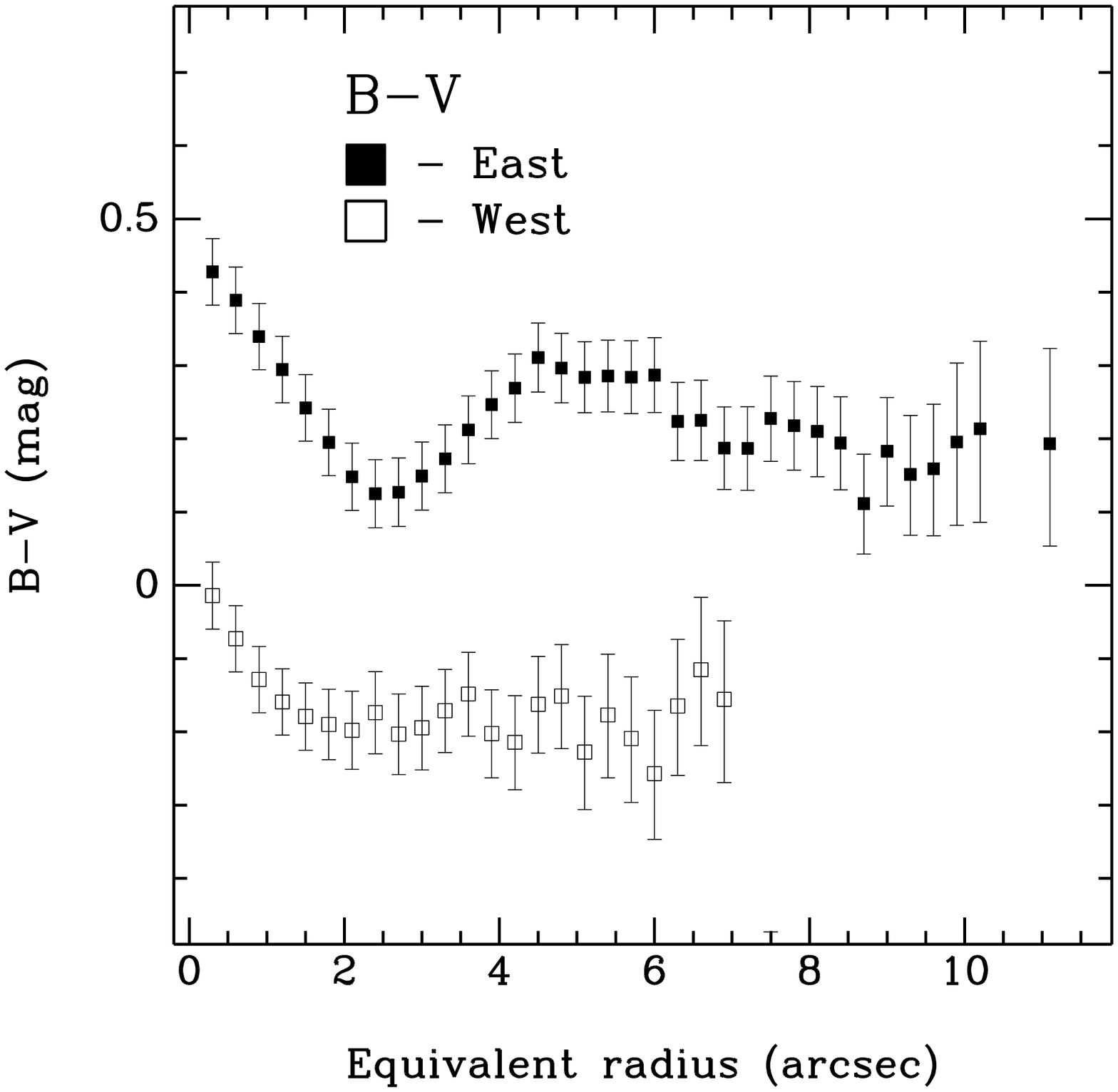}
   \includegraphics[width=5.9cm,bb=33 248 583 709,clip=]{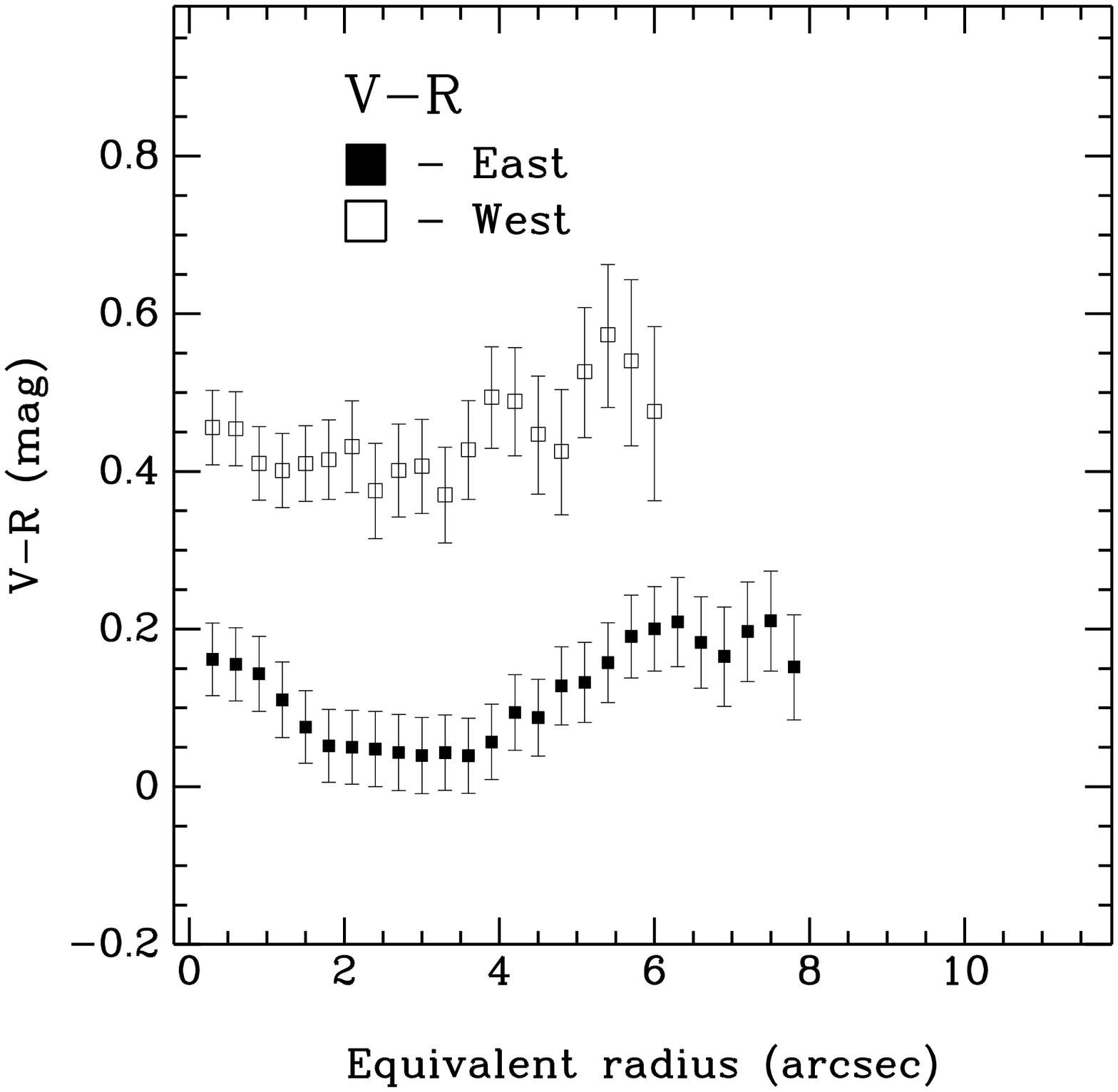}
   \includegraphics[width=5.9cm,bb=33 164 583 709,clip=]{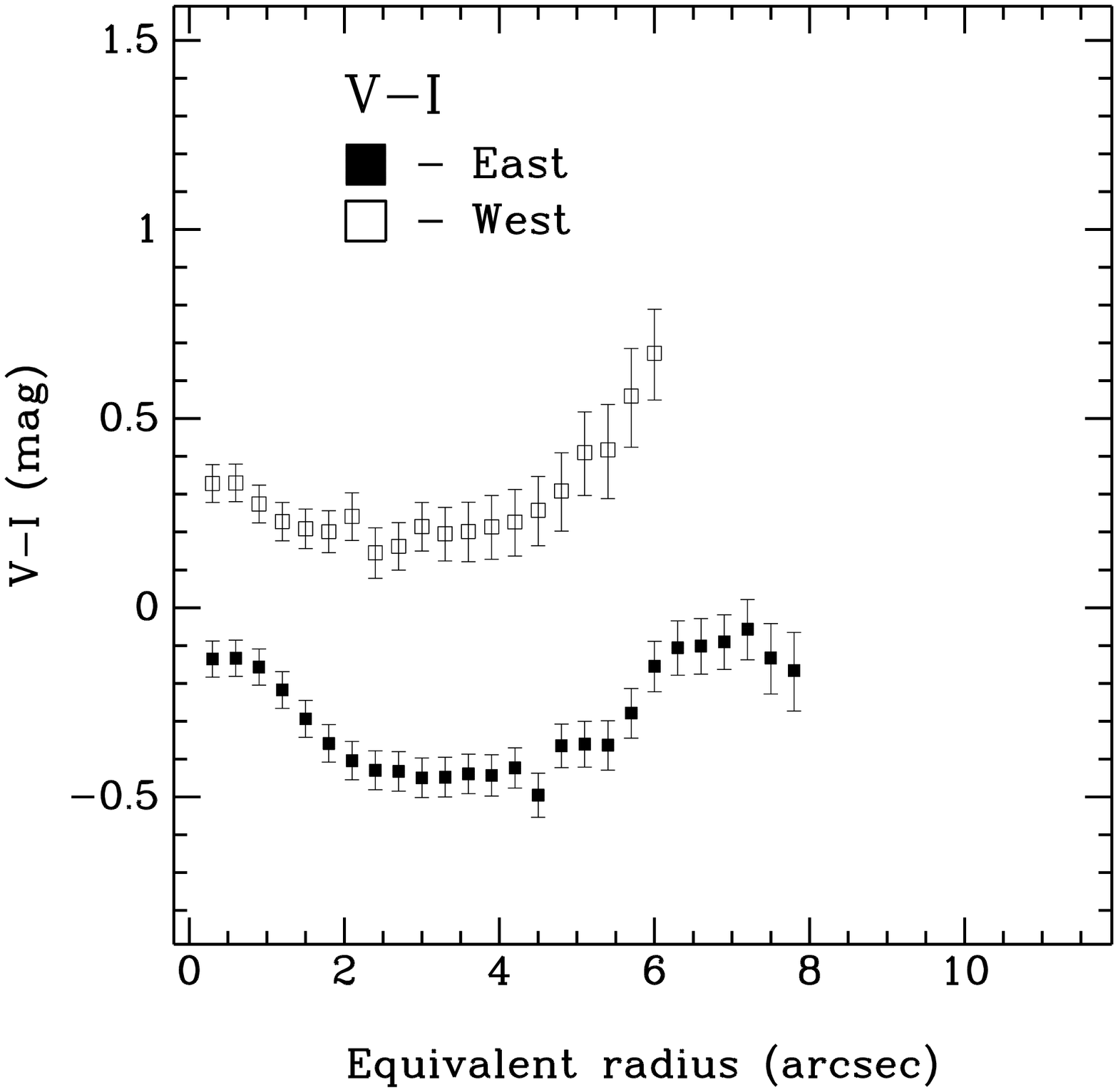}
   \includegraphics[width=5.9cm,bb=33 164 583 709,clip=]{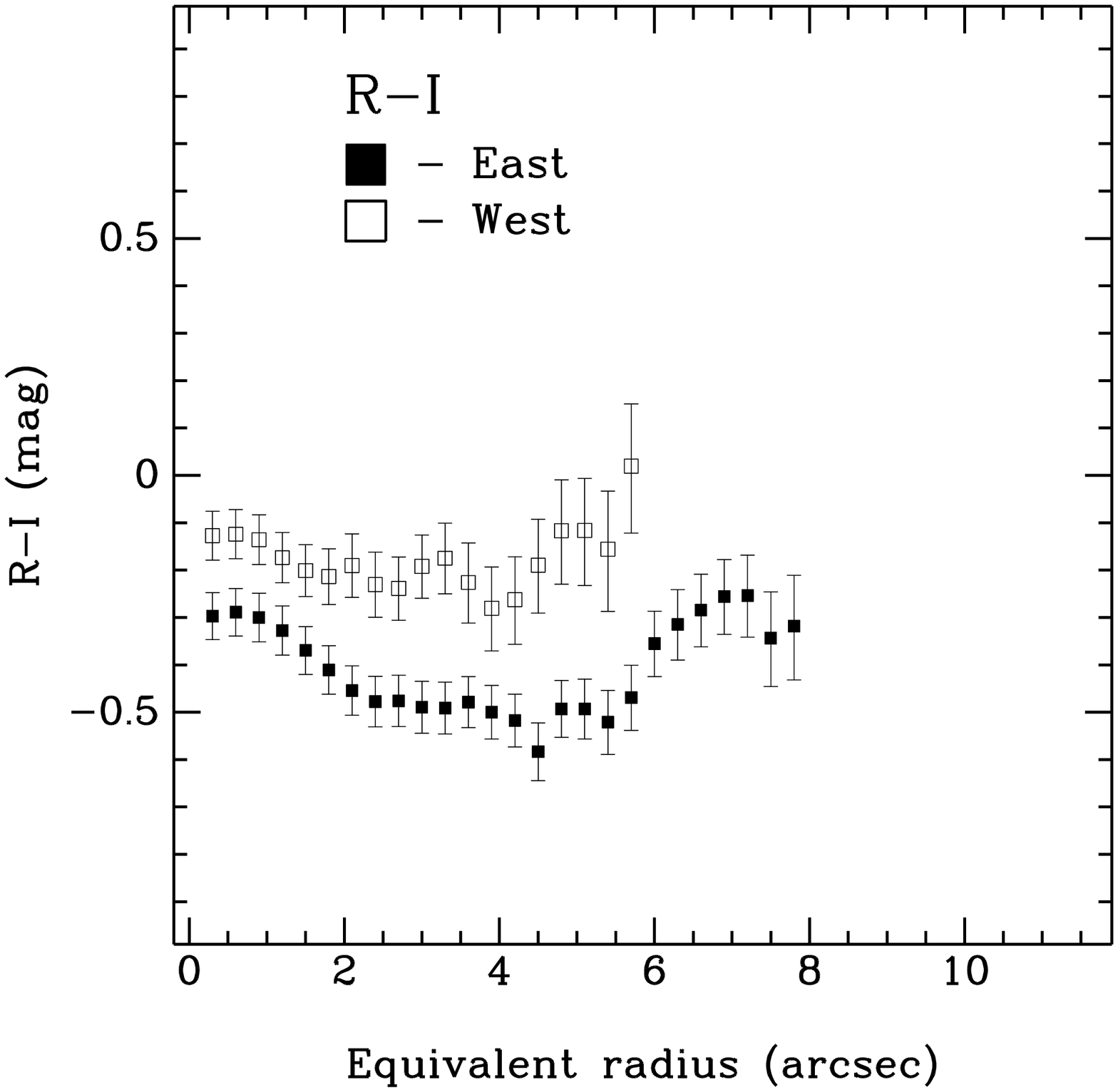}
   \includegraphics[width=5.9cm,bb=33 164 583 709,clip=]{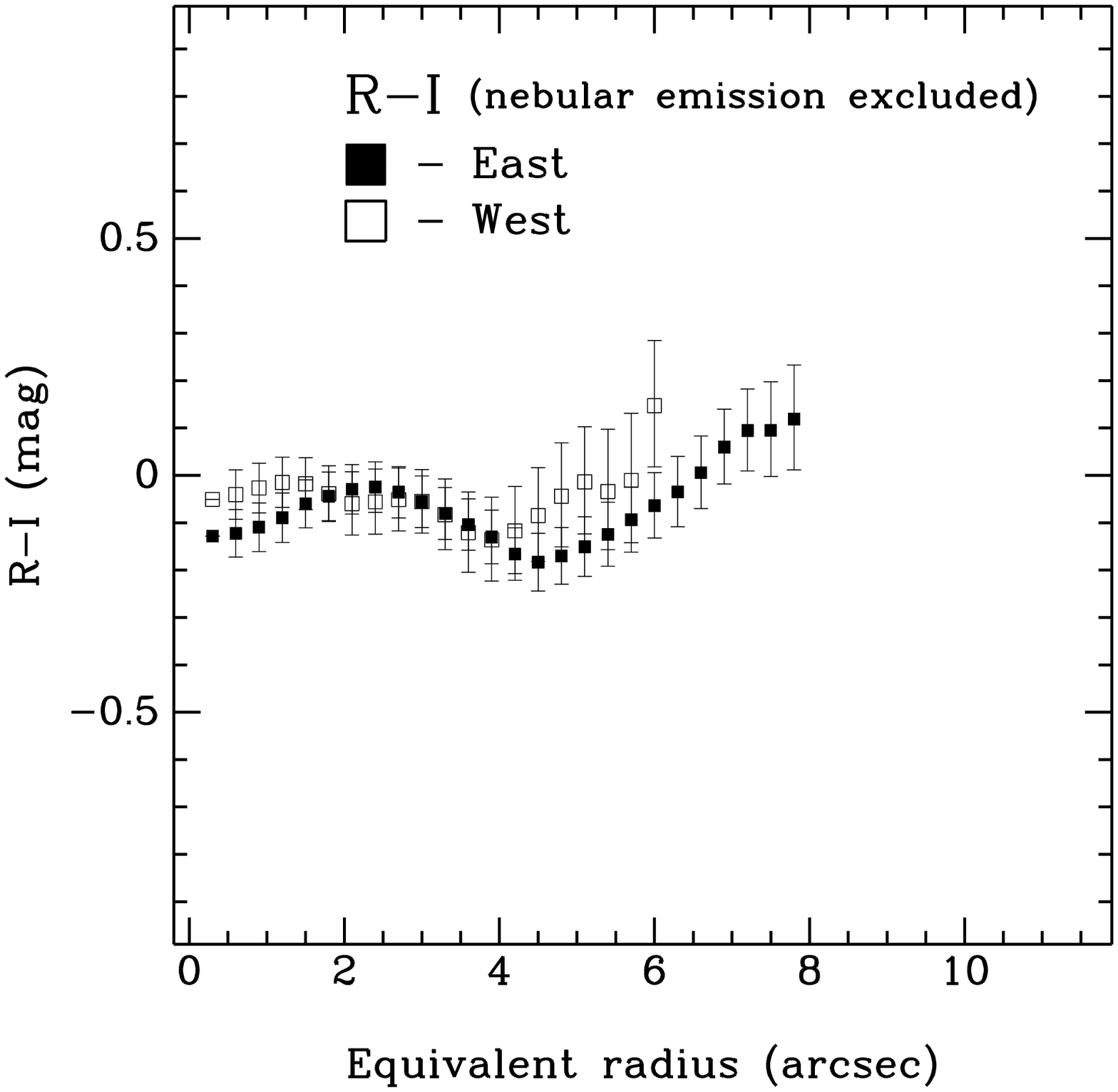}
      \caption{Radial distributions of colours for SBS 0335--052
	E and W,  corrected for Galactic extinction. Only
	$R-I$ gas-free colour distributions are shown since they
	are practically independent of the model assumptions
	for the ionized gas parameters (see Sect. \ref{method}).
              }
	 \label{FigColor}
   \end{figure*}

The $B$-band images of both galaxies are shown in Figure \ref{Image_B}
for further comparison with their images in the net H$\alpha$.
The results of our surface photometry are presented as the surface brightness
profiles (SBPs) in Figure \ref{FigSBP}. Each of the six panels shows SBPs of
both east and west galaxies in the respective broad-band and H$\alpha$
filters.
SBPs for the east and west galaxies are shown by filled and open squares,
respectively.
The H$\alpha$ SBPs were transformed from the instrumental magnitudes to some
conditional magnitudes using the same zero-point and colour term coefficients
as for the $R$-band.
This allows us to directly estimate the contribution of H$\alpha$ into
$R$-band flux from the plotted SBPs.
For example, at $R_{\rm eff}$=6\arcsec\ we have for the east galaxy
$\mu_{\rm R} \sim$24 mag~arcsec$^{-2}$, while $\mu_{\rm H\alpha} \sim$24.5
mag~arcsec$^{-2}$.
This implies that at this radius the flux in the H$\alpha$-line contributes
50\% of the total $R$-band flux (accounting for $R$-filter transmission of
$\sim$0.8 at the observed wavelength of H$\alpha$),
or the ratio of the line and continuum fluxes within $R$-filter band is
$\sim$1.0. Accounting for the equivalent width of the $R$-band of
$\sim$1600~\AA,
this means that at $R_{\rm eff}$=6\arcsec\  the
mean EW(H$\alpha$) is $\sim$1600~\AA. Similarly, at $R_{\rm eff}$=8\arcsec\
we can roughly estimate the mean EW(H$\alpha$) as $\sim$400~\AA. This
immediately shows the important effect of nebular emission on the colours of
the outermost LSB component.

The model fittings of the SBPs for the east and west galaxies are shown by the
solid and
dashed lines, respectively. We fitted their SBPs, assuming that the outer
parts
of both galaxies are well described by the exponential profile (kind of
disk): $\mu(r) = \mu_{\rm 0} + 1.086\times r/\alpha_{\rm 0}$.
The innermost bright region with active star formation was approximated
by a Gaussian profile
$\mu(r)= \mu_{\rm G} + 1.086\times\ln2\times(2\cdot r/\alpha_{\rm G})^{2}$,
where  $\alpha_{\rm G}$ is the FWHM of the Gaussian profile.
Since we are mainly
interested in the colours of the outermost parts of these galaxies, the exact
form of the SBPs in the central bright regions is outside the scope of this
study.
This is natural, accounting for the modest seeing we have for these data.

The derived model parameters with their uncertainties are given in
Table \ref{tabStrParam}. With the scalelengths of broad-band ``disk''
components of 1\farcs7--1\farcs9 for east, and 1\farcs6--1\farcs7 for
west galaxies, these disk-like profiles can be followed up to 4--5 scalelengths
($R_{\rm eff}$=10\arcsec\ and 7--8\arcsec, respectively) reaching the
surface brightness level of $\sim$27 mag.~arcsec$^{-2}$ in $B$ and $V$,
and about 1 mag brighter in $U$ and $I$.
The H$\alpha$ SBPs can also be followed in the outer parts with the same
$R_{\rm eff}$, and are more or less well described outside the bright
star-forming regions by an exponential law.
Since the seeing FWHM is comparable to the `disk' scalelength, we have
modeled the convolution of the found disk with its respective Gaussian to
check
how much the seeing affects the `disk' parameters. The effect is small:
$\mu_{0}$ gets fainter by 0.05 mag, while the scalelength increases by 2\%.
The effect of seeing on the parameters of the central Gaussian is very large:
its central brightness drops by 1.15 mag, and the FWHM increases by a factor
of 1.6. Therefore we do not discuss the central Gaussian parameters below.
Disk fitting for the W galaxy is further complicated by the presence of
two distinct knots with a distance of $\sim$4\arcsec. Since the
maximum effective radius to which the light of the W galaxy is followed is
only $\sim$6--8\arcsec, in some cases (in $U$) it is difficult to
perform a reasonable exponential fitting to the SBP.

In columns 7 and 8 of Table \ref{tabStrParam} we also present the total
magnitudes of the LSB `disk' components and those of the starburst components.
The former is calculated by the standard formula for an exponential disk
using the parameters $\mu_{0}$ and $\alpha_{0}$:
$$ m_{\rm disk} = \mu_{0} - 5.0\times\log(\alpha_{0}) - 2.5\times\log(2\pi)$$
The magnitudes of the starburst component are calculated from the total
magnitudes of the galaxy in Table \ref{tabTotMag} subtracting the flux of
the LSB `disk'.

It is worth noting that the scalelengths of `disk' components are quite close
for the east and west galaxies (mean $\sim$1\farcs78 and 1\farcs70,
respectively),
while the central SB $\mu_{\rm B,corr}^{0}$ differ significantly. For the east
galaxy this parameter is 21.07 mag~arcsec$^{-2}$, typical of BCGs.
For the west galaxy this parameter is 22.29 mag~arcsec$^{-2}$,  more
typical
of dIrr galaxies. Furthermore, the apparent strength of the current SF episode
differs drastically in these two galaxies, despite the fact that their ages
are rather close,
$\sim$3--3.8 Myr (see Sect. \ref{SFR}). From the numbers in
columns 7 and 8 of Table \ref{tabStrParam} one can estimate that the relative
$B$-band luminosity of the starburst ($L_{\rm burst}$/$L_{\rm disk}$) is
lower in the west galaxy compared to that of the east galaxy by a factor of
$\sim$8.
The size of the bright `Gaussian' component decreases systematically from
$U$ to $I$ for both galaxies, that mainly reflects the wavelength
dependence of the seeing during observations.
However, there is also a systematic difference of this size between the east
and
west galaxies for each band. This indicates
that the brightest knot of the west galaxy is more compact than that for
the east galaxy.

In Figure~\ref{FigColor}  we present the radial distributions of colours
in both objects. Again, the filled squares are used for the east galaxy,
and the open ones for the west galaxy. Since the seeing for the $U$-band
image
is significantly larger than that for $B$-band, to derive the $(U-B)$ radial
profile, we convolved the $B$ image to the effective seeing of the $U$-band
image
(that is obtain the resulting FWHM for surrounding stars equal to that on
the $U$ image).

\subsection{H$\alpha$ map and equivalent width}

Since our main goal is to derive in both galaxies the true colours of
the LSB underlying components, comprised of an older stellar population, we
need to examine the contribution of the ionized gas emission so as to
subtract it properly. The effect of this contribution can
be followed directly from observations, e.g., from the map of the value of
EW(H$\alpha$).
Since the equivalent width of the $R$-band filter is $\sim$1600~\AA, an
additional gas emission with EW(H$\alpha$) $\gtrsim$ 300--400~\AA\ will
result in a significant shift of the original colours of the underlying
stellar
population. For purely ionized hydrogen emission with $T_{\rm e} \sim$20000 K
the equivalent width EW(H$\alpha$)$\sim$3500~\AA\ (e.g., Aller \cite{Aller84}).
So, even diluted radiation of the ionized gas
can significantly alter the intrinsic colours of underlying stars.

\begin{figure*}
   \centering
   \includegraphics[width=6.8cm,bb=28 248 583 719,clip=]{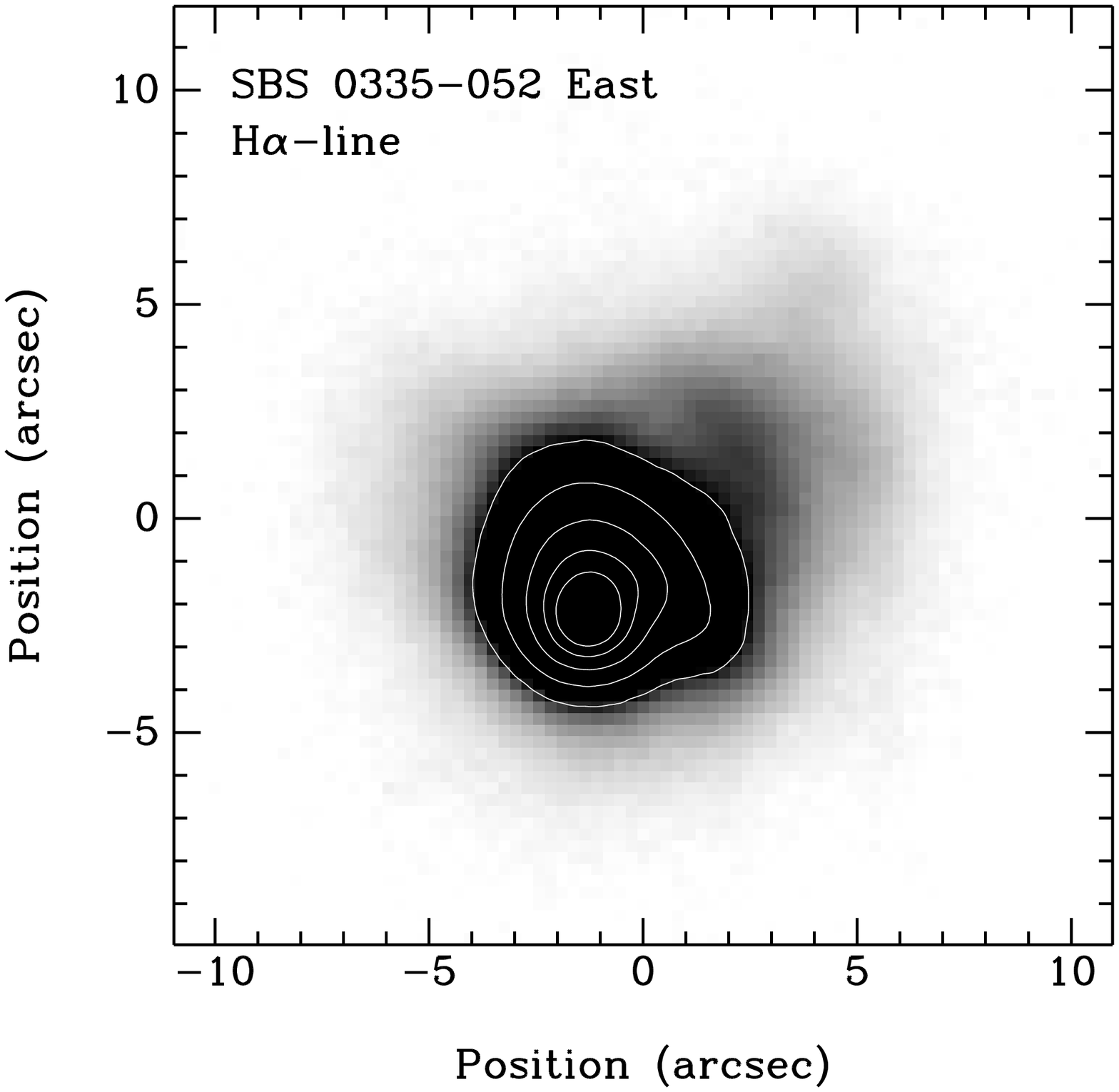}
   \includegraphics[width=6.8cm,bb=28 248 583 719,clip=]{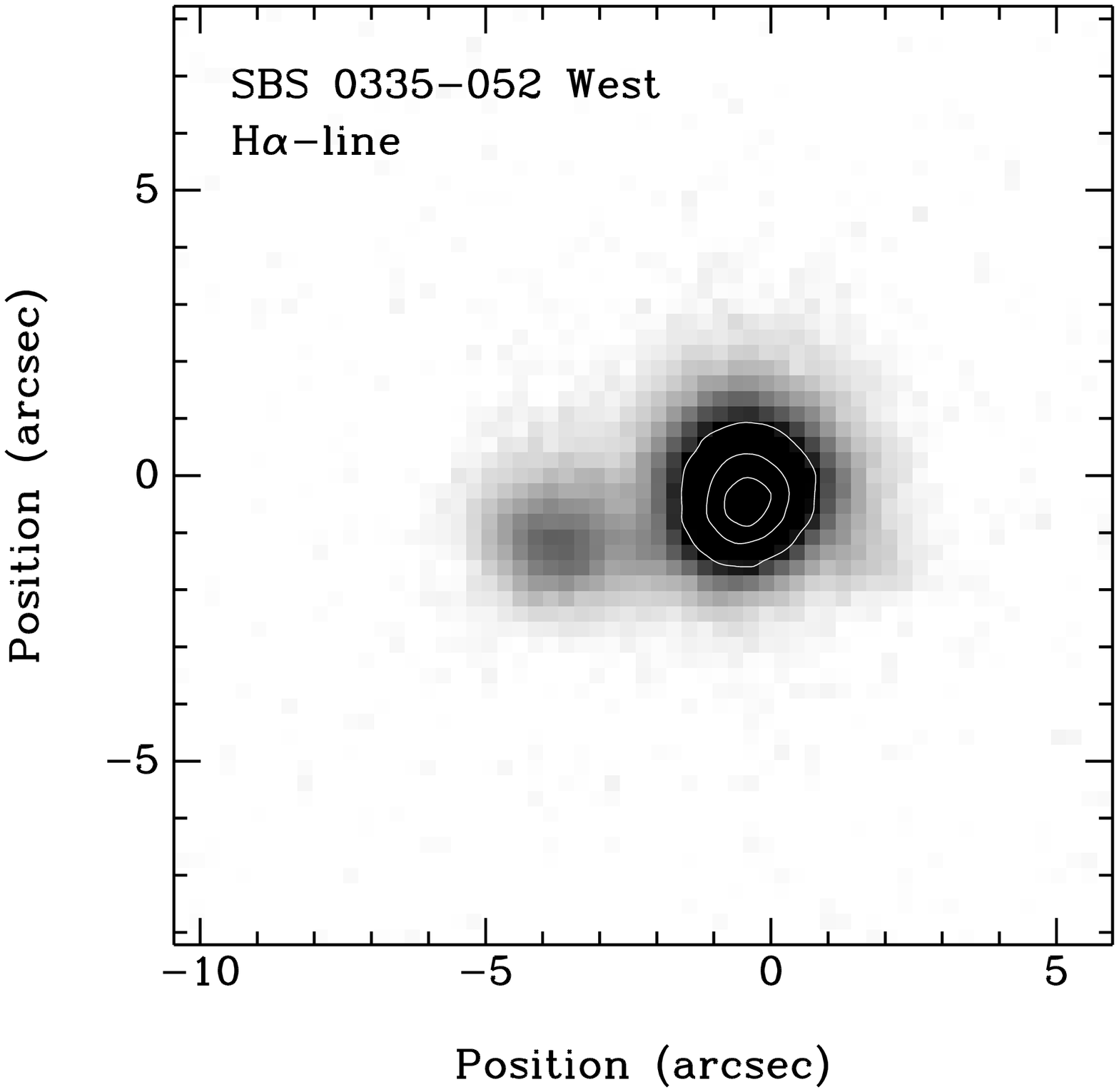}
   \includegraphics[width=6.8cm,bb=28 248 583 709,clip=]{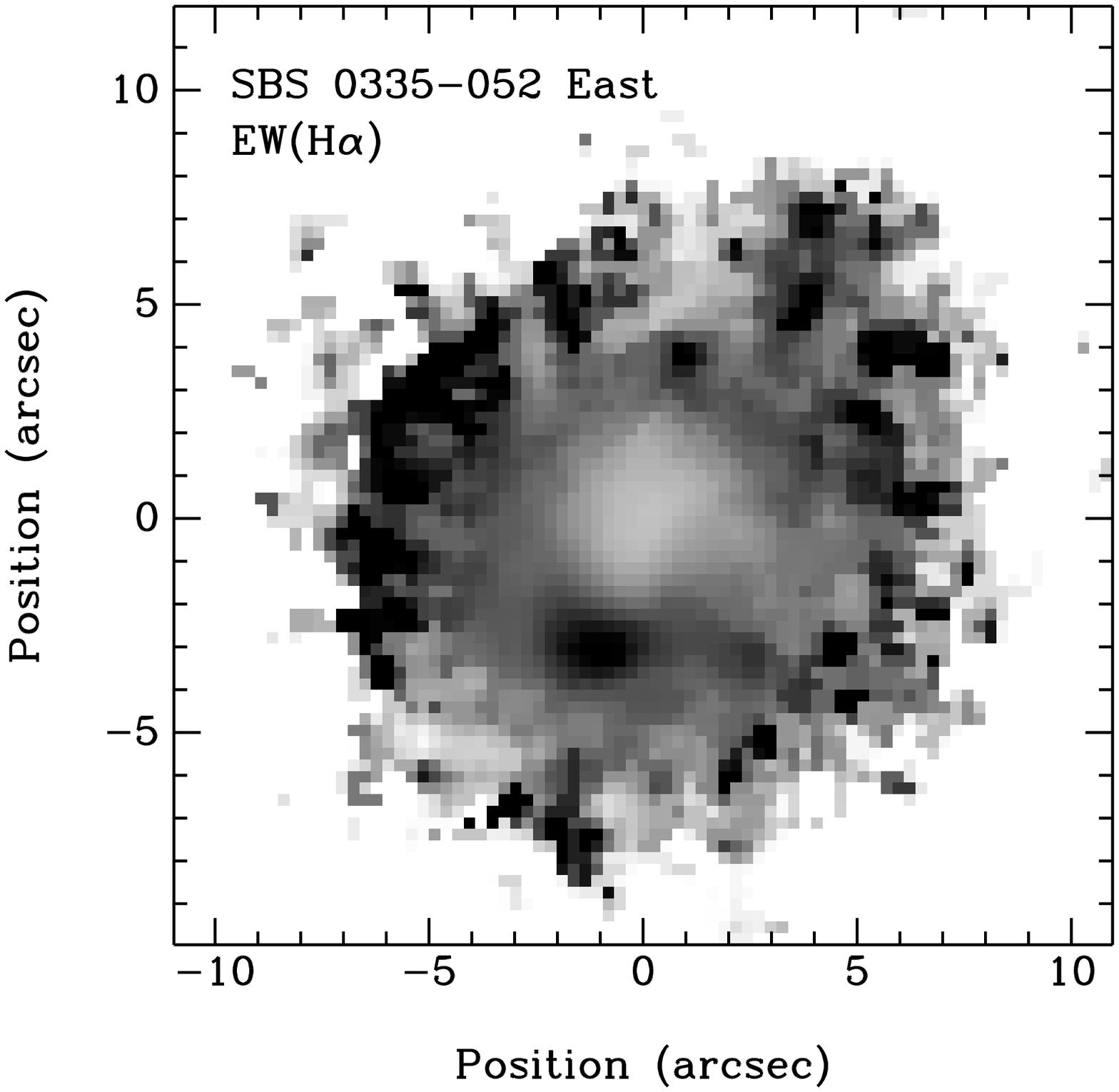}
   \includegraphics[width=6.8cm,bb=28 248 583 709,clip=]{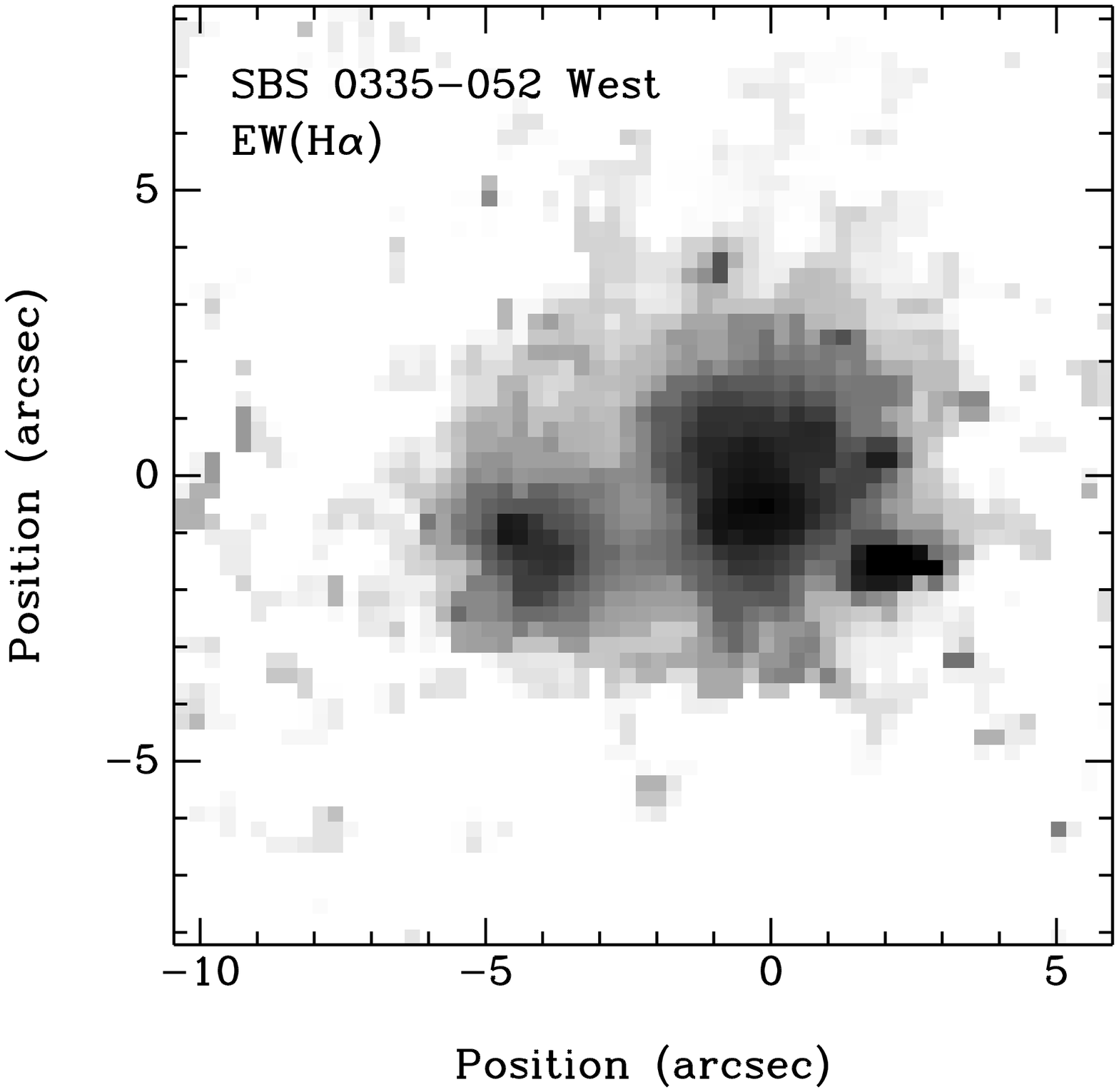}
   \includegraphics[width=6.8cm,bb=28 164 583 709,clip=]{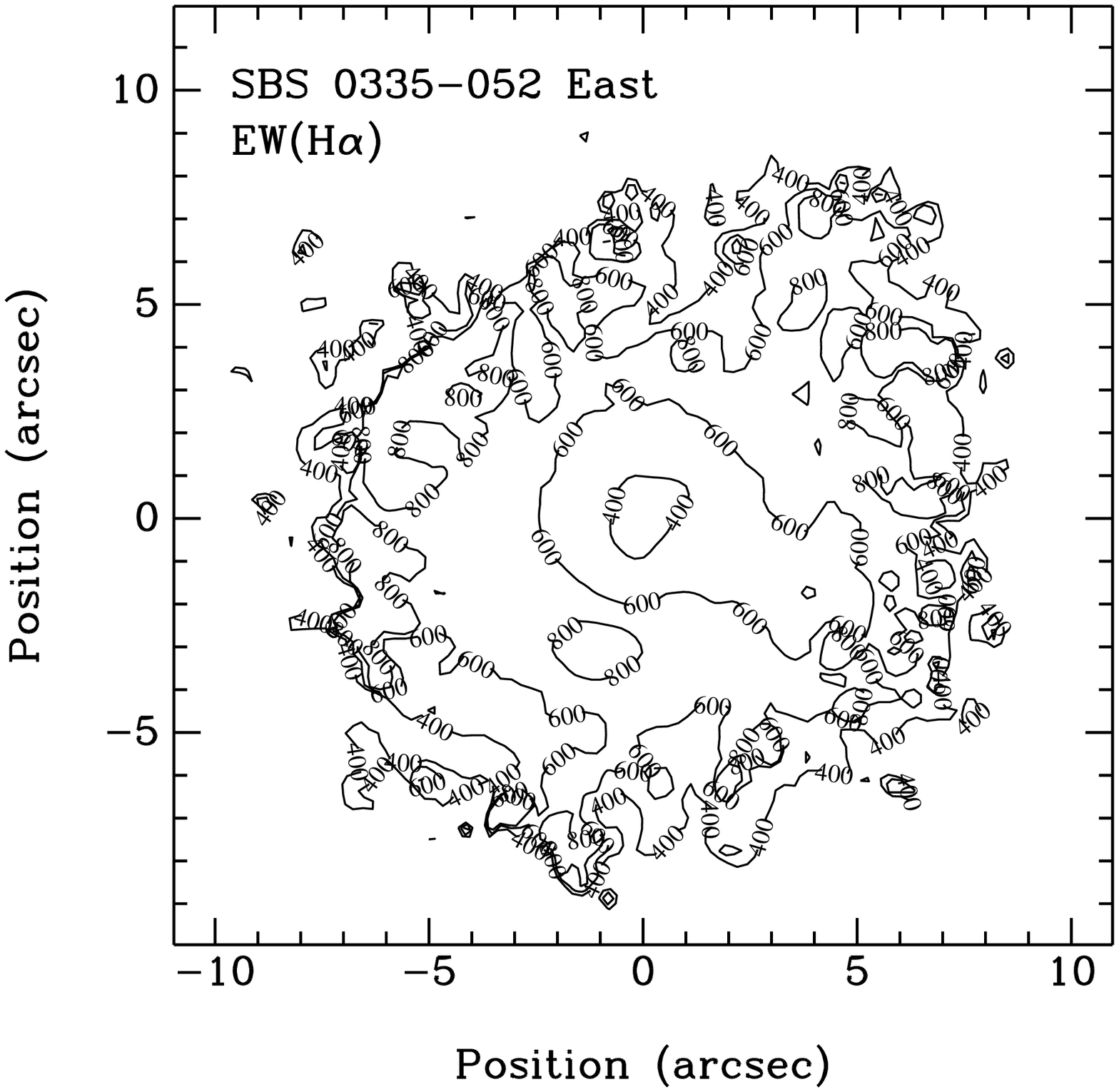}
   \includegraphics[width=6.8cm,bb=28 164 583 709,clip=]{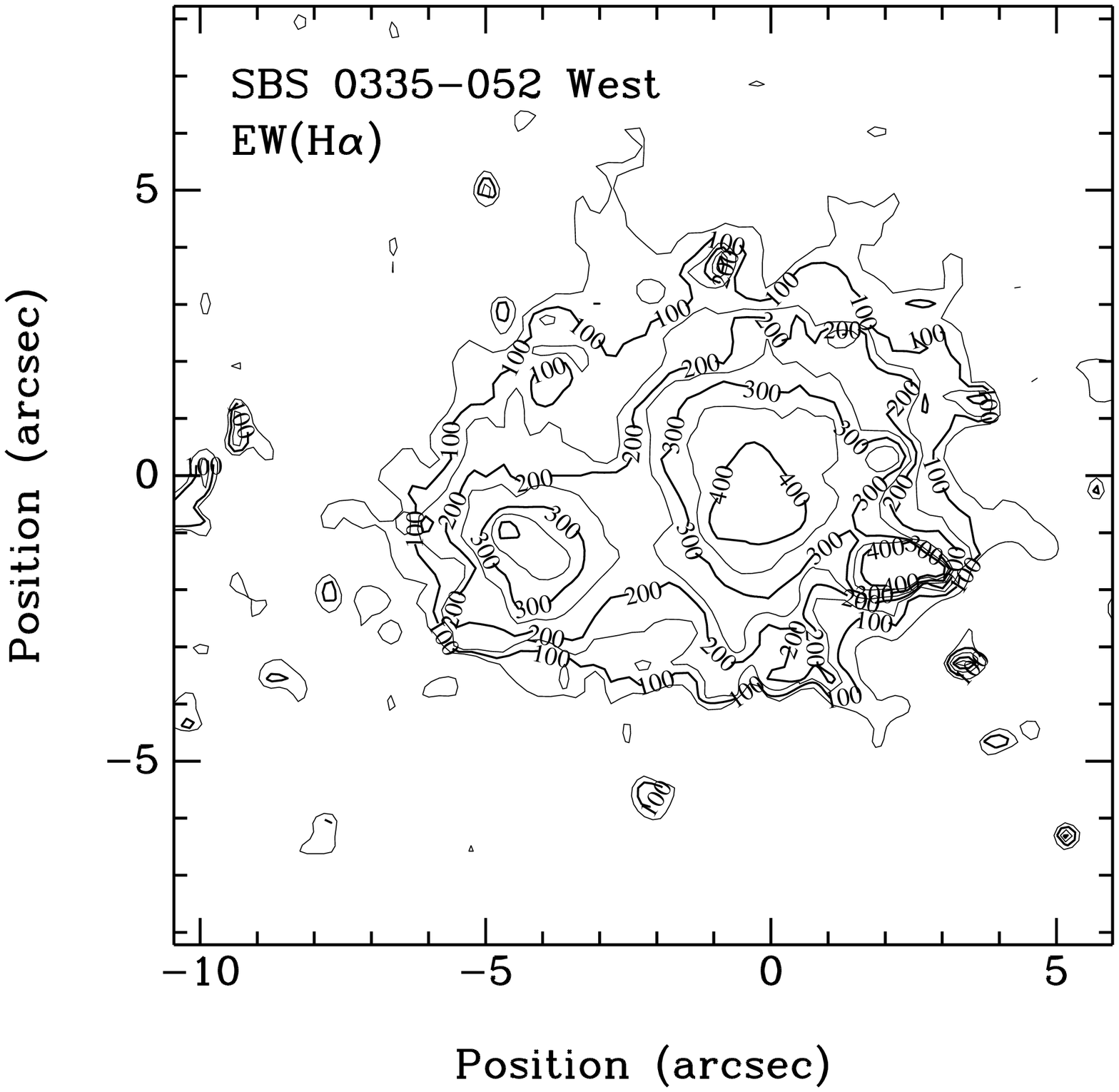}
      \caption{From top to bottom: images of H$\alpha$-line flux, grey-scale
	       and contour maps of EW(H$\alpha$) for galaxies
	       SBS 0335--052 E and W. On each image grey-scale is
	       normalized to the respective maximum value.
	       Isophotes superimposed on H$\alpha$ flux image correspond
	       to the surface brightnesses  20.3, 20.7, 21.5, 22.5 and 23.5
	       mag~arcsec$^{-2}$ -- for the east galaxy, and 23.2, 23.5 and 24.1
	       mag~arcsec$^{-2}$ -- for the west galaxy.
              }
	 \label{FigHalpha}
   \end{figure*}

To illustrate the importance of ionized gas emission, we show in
Fig.~\ref{FigHalpha} the images of both galaxies in the net H$\alpha$-line
(with the line-free continuum subtracted) and the maps of distribution of
EW(H$\alpha$), both in grey-scale and as isolines with the values of EW drawn.
One can see that in the east galaxy the EWs of H$\alpha$ in the outermost
regions reach the level of 400--800~\AA. In the west galaxy they are in general
lower, and have the values of 100--200~\AA. These diluted values of EWs show
the relative contribution of the ionized gas emission and the stellar light
in these regions to the $R$-band.
Accounting for the $R$-filter transmission (0.795) at H$\alpha$ with the
given galaxy redshift and the effective width of $R$-filter of $\sim$1600~\AA,
we estimate that
in the east galaxy the nebular emission (H$\alpha$ + continuum) will
brighten the stellar continuum by $\sim$0\fm33--0\fm65.
In the west galaxy this brightening is $\sim$0\fm08--0\fm16.

\section{Models and evolutionary status}
\label{models}

\subsection{Method description}
\label{method}

The observed  light from an actively star-forming galaxy is a mix of
the stellar and ionized gas radiation affected by the dust extinction
inside the galaxy and in the Milky Way. The maximum portion of nebular
emission
emerges from the giant \ion{H}{ii} regions surrounding young stellar clusters
often situated close to the central parts of star-forming galaxies. However,
nebular emission can also significantly contribute to the light of the
outer parts of some of blue compact galaxies.
For this reason, in the analysis of stellar ages we can not directly use the
results of the broad-band surface photometry. Instead, we need to take into
account the possible contribution of nebular emission. To a first
approximation,
this can be done based on the intensity of H$\alpha$ emission over the
galaxy body and on the information on the radial trends of such parameters
as electronic temperature and element abundances,
derived from the long-slit spectroscopy results
(here, from Izotov et al. \cite{Izotov99}, \cite{Izotov01}).

Thus, one of the necessary steps is the creation of the ionized gas nebular
spectrum for each studied position of the galaxy image. This should be
subtracted from the original broad-band fluxes at each point. This nebular
spectrum consists of two parts. The first one is a purely recombination
spectrum of hydrogen-helium plasma described, e.g., in Aller
(\cite{Aller84}).
It includes lines of hydrogen, \ion{He}{i} and \ion{He}{ii} (with the latter
derived from the accepted ratios of $N(\ion{He}{^+})/N(\ion{H}{^+})=0.078$ and
$N(\ion{He}{^{++}})/N(\ion{H}{^+})=0.003$ (Aller \cite{Aller84};
Izotov \cite{Izotov99}). The continuum includes hydrogen free-free,
free-bound and the two-photon radiation, and the respective components from
\ion{He}{i} and
\ion{He}{ii}. This contribution can be estimated accurately,
given the
observed flux in the H$\alpha$-line, if we know the ionized gas electron
temperature T$_{\rm e}$ distribution.

The second part of the nebular emission is related to the contribution of the
strong emission lines of oxygen [\ion{O}{ii}]~$\lambda$3727, [\ion{O}{iii}]
$\lambda\lambda$4959,5007 and moderate lines of other ions (e.g.,
[\ion{Ne}{iii}]~$\lambda$3868). This is more uncertain, since it depends on
the possible small variations in element abundances and the gas temperature.
The easiest way to account for nebular emission in these galaxies is to use
the results of their long-slit spectroscopy for the maximal radial distances
from the bright starburst. Unfortunately, such data are available for
the east galaxy only for two position angles and for the west galaxy for one
position angle (Izotov et al. \cite{ILC97, Izotov99, Izotov01}; Lipovetsky
et al. \cite{Lipovetsky99}). Since the
hypothesis of the circular symmetry of these galaxies is not too realistic,
these long-slit results can be used only as a first approximation.

Since we are interested in the colours of the underlying low-surface
brightness components, we examined the effect of nebular emission
only for the outermost parts of  both galaxies, where we can
neglect the scattered light of bright starbursts. However,
to have a sufficiently large signal-to-noise ratio in the studied
regions, we are limited by the effective radii, which correspond to
$\mu_{\rm B} \lesssim$ 25.5 mag~arcsec$^{-1}$ for the east galaxy
($R_{\rm eff}\approx$7.5\arcsec) and 25.2 mag~arcsec$^{-1}$ for the west
one ($R_{\rm eff}\approx$5\arcsec).

To minimize the effect of different seeings in various broad-band
filters and in H$\alpha$ on the derivation of the gas-free colours, we
convolved the H$\alpha$ image to the seeing of the respective broad-band
image.
This was performed before subtracting nebular the emission from $U$, $B$ and
$V$
images. For $R$ and $I$ images the seeing was the same as for the H$\alpha$
filter.
However, due to the large difference between the seeing in the $U$-band image
on the one hand, and in $B$-band and H$\alpha$-filter images, on the other
hand, some systematic effects in the colour $(U-B)$ probably diminish
the accuracy of our estimate of its gas-free part.

In our calculations of the nebular emission contribution we accepted
the relative intensities of emission lines at respective radial distances
(see Table~\ref{tabGasInt}) as derived from the
2D long-slit Keck telescope spectrum, kindly provided by Y.Izotov (Izotov et
al. \cite{Izotov01}).
To account for the possible variations of the relative intensities on azimuth,
we allowed that the average line intensities of strong lines in the ring
can differ from those measured in some specific sector of the ring by as much
as 5\%.
This is consistent with the maximal variations along the slit seen in the
periphery of this galaxy.
This uncertainty can
result in additional errors of the gas-free colours of the level of 0.03
-- 0.04 mag.

Thus, the estimates of the gas-free colours
have been made on the regions with $R_{\rm eff}\approx$7.5\arcsec\
for the East galaxy and with $R_{\rm eff}\approx$5\arcsec\
for the West galaxy.
To investigate the influence of gas electron temperature on the estimated
stellar ages, we created two versions of the ionized gas spectra with
$T_{\rm e}$=15000~K and 20000~K. As shown by Izotov et al.
(\cite{Izotov99,Izotov01}), the electron temperature within the supergiant
\ion{H}{ii} region of SBS~0335-052~E has no significant gradient and
remains in the range of $T_{\rm e}$ from 18000 to 23000~K.
For the considered nebular emission, the electron temperature
$T_{\rm e}$ was adopted to be of
20000~K. If it would be as low as 15000~K, the derived colours of the
gas-free `disk' would be at most 0.02 mag. bluer.

For each  pixel we connected
the intensities of the model nebular spectra (lines and continuum) with
the observed H$\alpha$ flux in the pixel, according to the relations from
Aller (\cite{Aller84}).
Then the model nebular spectra were redshifted to $z$=0.0134,
and convolved  with the passband transmission curves
of the $U, B, V, R$ and $I$ filters. The derived flux of nebular emission in
each of the filters was subtracted from the fluxes of the respective
original CCD images, corrected for Galactic extinction according to the
Whitford (\cite{Whitford58}) law, pixel by pixel.
For the resulting CCD images with subtracted nebular emission
we built in each filter the SBPs of the gas-free light.

\begin{table}[hbtp]
\caption{Adopted line intensities of the model spectra}    
\centering{
\label{tabGasInt}
\begin{tabular}{lclc} \hline \hline
\rule{0pt}{10pt}
\rule{0pt}{10pt}
$\lambda_{0}$(\AA) Ion &I($\lambda$)/I(H$\beta$)  & $\lambda_{0}$(\AA) Ion &I($\lambda$)/I(H$\beta$)  \\ \hline
3727\ [\ion{O}{ii}]\    & 0.43 & 4363\ [\ion{O}{iii}]\  & 0.05   \\
3750\ H12\              & 0.03 & 4471\ \ion{He}{i}      & 0.04   \\
3771\ H11\              & 0.04 & 4686\ \ion{He}{ii}     & 0.03   \\
3798\ H10\              & 0.05 & 4861\ H$\beta$\        & 1.00   \\
3835\ H9\               & 0.07 & 4959\ [\ion{O}{iii}]\  & 0.50   \\
3868\ [\ion{Ne}{iii}]\  & 0.30 & 5007\ [\ion{O}{iii}]\  & 1.50   \\
3889\ H8\               & 0.11 & 5876\ \ion{He}{i}      & 0.06   \\
3967\ H7\               & 0.25 & 6563\ H$\alpha$\       & 2.73   \\
4101\ H$\delta$\        & 0.26 & 6678\ \ion{He}{i}      & 0.05   \\
4340\ H$\gamma$\        & 0.48 & 7065\ \ion{He}{i}      & 0.05   \\
\hline \hline
\end{tabular}
}
\end{table}

The derived gas-free colours $(U-B), (B-V), (V-R)$ and $(R-I)$ at the
respective
$R_{\rm eff}$, corrected for the Galactic extinction,
are summarized for both galaxies in the 3-d column of Table \ref{tabStarAge}.
The radial distributions of gas-free colour $(R-I)$, as the least
model-dependent one, are shown in Fig. \ref{FigColor} (bottom-right panel).
This gas-free colour is the most robust, since the nebular emission due
to the strong H$\alpha$ emission  in the $R$-band is determined directly from
observations,
while in $I$-band the nebular continuum and He lines are also determined
directly and the strong lines of other elements are absent. Possible
variations in the fluxes of \ion{He}{i}~$\lambda$6678 and
[\ion{S}{ii}]~$\lambda$6716,6731 lines affect the $R$ and $I$-band magnitudes
by less than 0\fm02.
While the original $(R-I)$ colours differ significantly, the gas-free ones
are very close, and show only mild variations with radius, between $-0.1$ and
0.0 -- for the W galaxy, and between $-0.1$ and +0.1 -- for the E galaxy.

Since there is no reliable information on the dust distribution in the
outermost parts of both the E and W galaxies, we did not include this in
the procedure of gas-free colour estimates. If some small internal dust
extinction is present ($E(B-V) \lesssim$ 0.20, Izotov et al. \cite{Izotov01}),
its effect on the gas-free colours will depend on the mutual distribution of
`old' stars and the ionized gas. If the $z$ scale for both components is
close,
they will be well mixed within the same volume, and the full effect of the
extinction will be the same for gas and stars. In this case it will be
equivalent to the additional Galactic extinction, and thus will have
no effect on the resulting colours of the stellar population. If the ionized
gas `disk' is thicker than the stellar one,  the value of the reddening of the
stellar light will be somewhat higher than that of the gas. This difference,
if present, will lead to bluer extinction-corrected colours and resulting
younger ages of underlying stars.
Since we ignore the possible internal extinction, the derived ages of old
stellar populations are, in this aspect, also the upper limits of the real
ages.

\subsection{Models to estimate the ages of underlying populations}
\label{ages}

The derived $(U-B), (B-V), (V-R), (V-I)$ and $(R-I)$ colours of the underlying
stellar population have been compared with the respective model colours from
the PEGASE.2 package (Fioc \& Rocca-Volmerange \cite{Pegase2}).
PEGASE.2 supersedes the previous
version of the spectrophotometric evolution model for starbursts and evolved
galaxies of the Hubble sequence (Fioc \& Rocca-Volmerange \cite{Pegase}).
The main differences of the new version are stellar evolutionary tracks with
non-solar metallicities, the library of stellar spectra
of Lejeune et al. (\cite{Lejeune97}, \cite{Lejeune98}) and
radiative transfer computations to model the extinction.
The codes of PEGASE.2 take into account the evolutionary
tracks of Girardi et al. (\cite{Girardi96}), Fagotto et al.
(\cite{Fagotto94a}, \cite{Fagotto94b}, \cite{Fagotto94c}) and
Bressan et al. (\cite{Bressan93}).
The models of Groenewegen \& de Jong (\cite{Groenewegen93})
are used to compute the AGB and post-AGB phases.

\begin{figure*}
   \centering
   \includegraphics[width=5.9cm,bb=33 164 583 709,clip=]{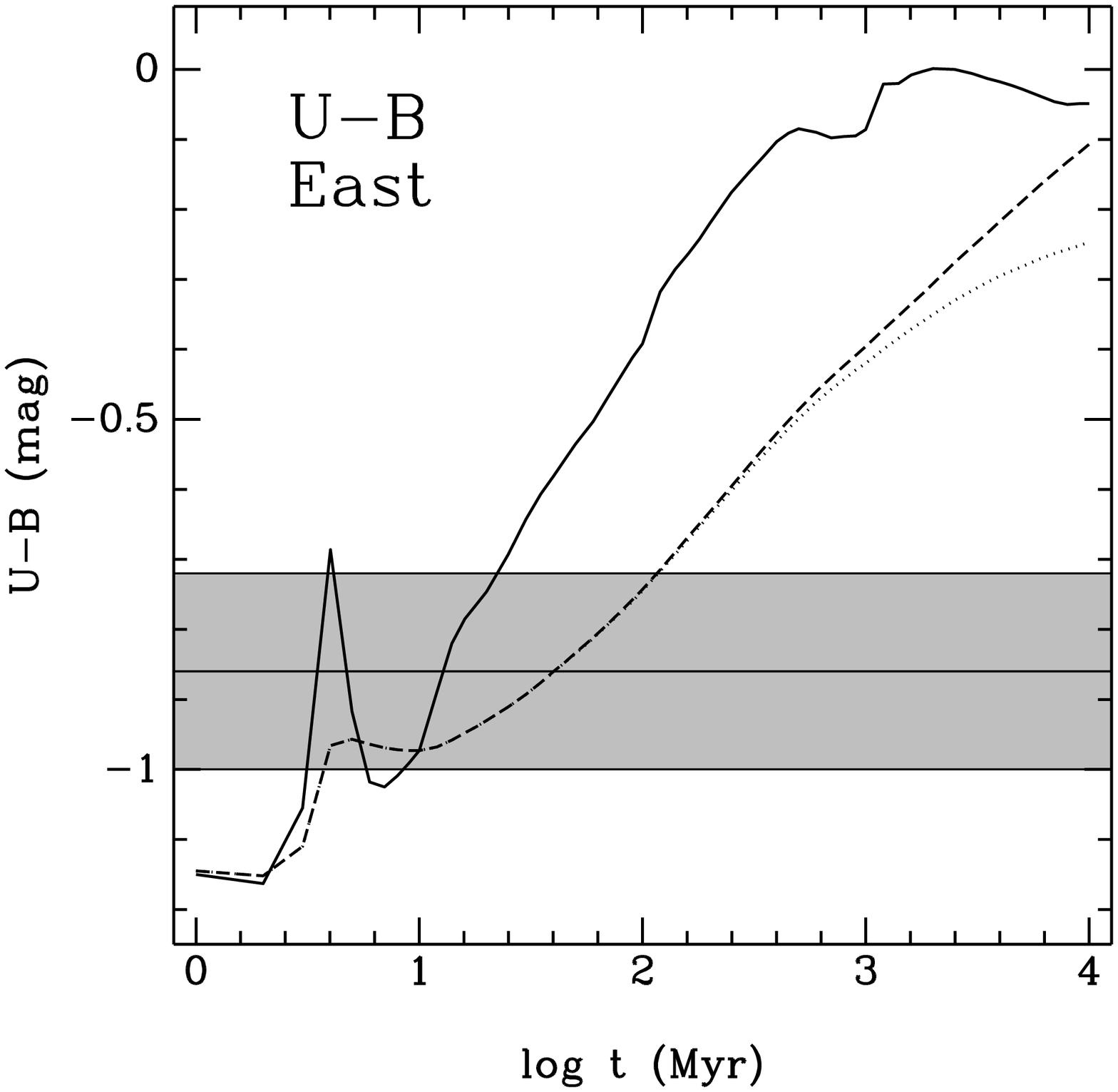}
   \includegraphics[width=5.9cm,bb=33 164 583 709,clip=]{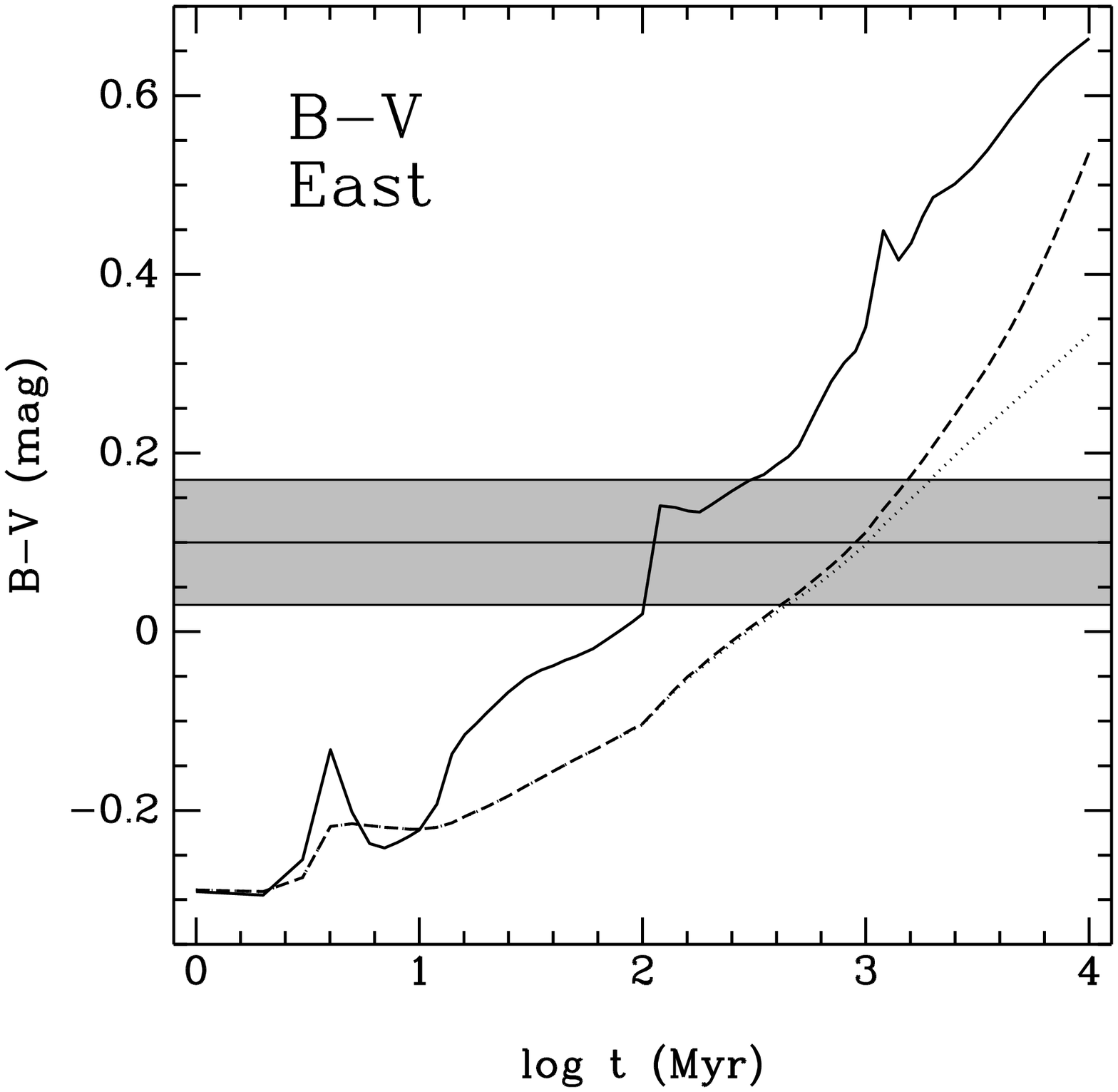}
   \includegraphics[width=5.9cm,bb=33 164 583 709,clip=]{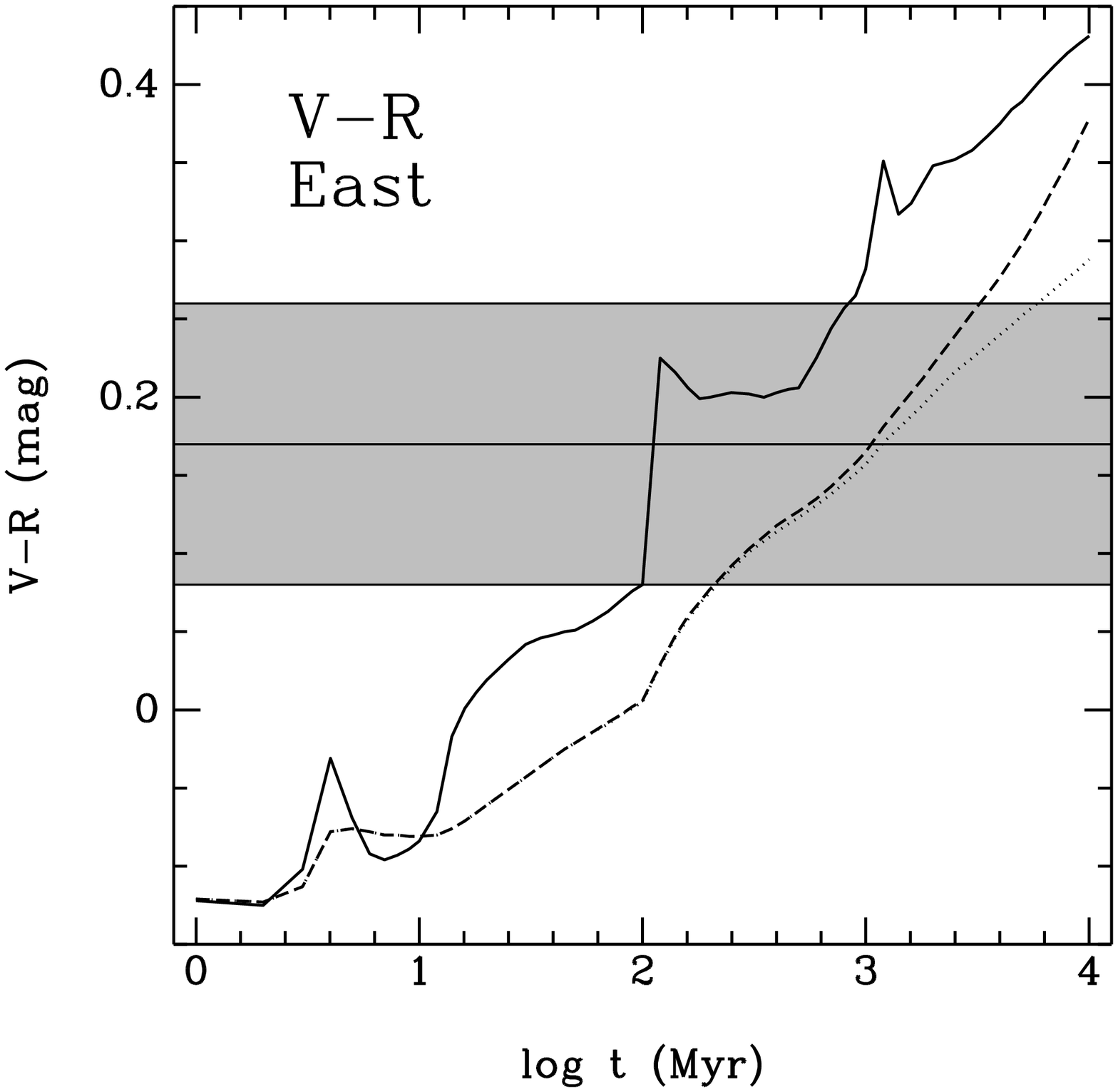}
   \includegraphics[width=5.9cm,bb=33 164 583 709,clip=]{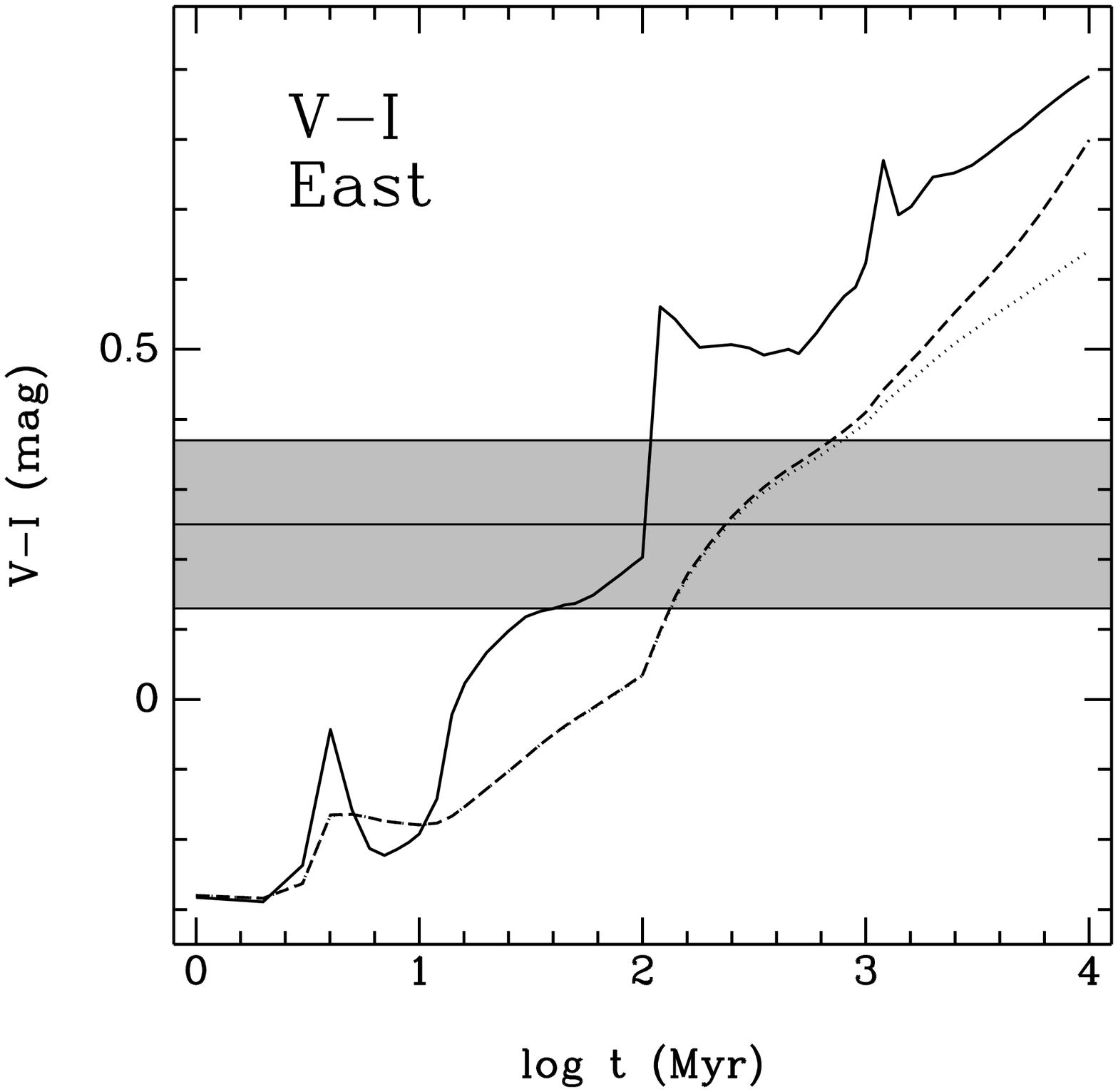}
   \includegraphics[width=5.9cm,bb=33 164 583 709,clip=]{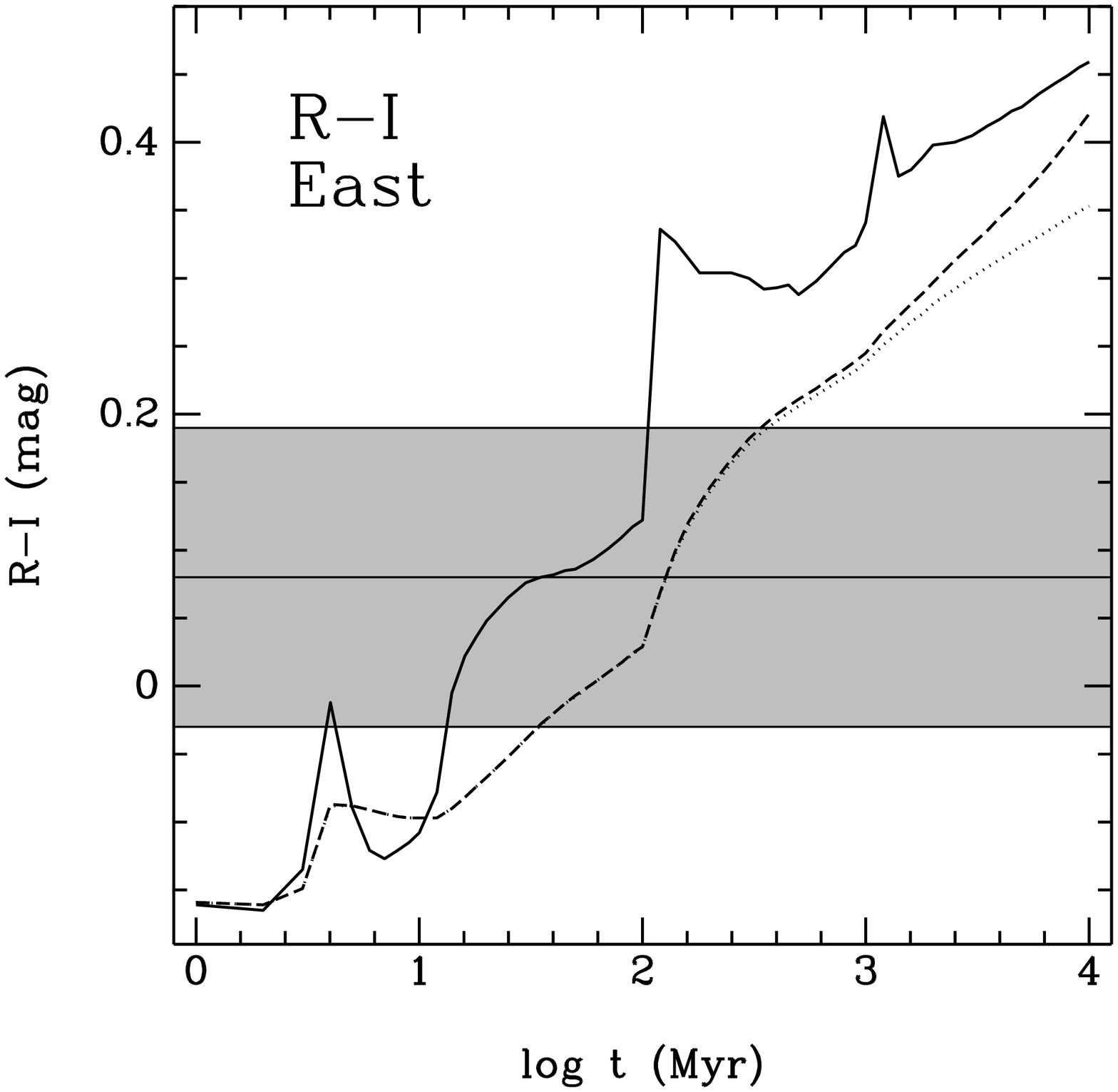}
      \caption{PEGASE.2 model evolution tracks (see text for details).
	 Solid line -- instantaneous burst model,
	 dotted line -- constant star formation rate,
	 dashed line -- exponentially decreasing
	       star formation rate with $\tau$=3~Gyr.
	 Horizontal lines show derived colours of the underlying LSB component
	 of the east galaxy with nebular emission subtracted. Shadowed regions
	 give the ranges of $\pm$1~$\sigma$ uncertainties of the derived
	 colours.     }
	 \label{FigModel_E}
   \end{figure*}

For calculation of the model colours,
we accept the metallicity $Z$\sunn/50 (the nearest of the available model
parameters to the observed ones), the Salpeter (\cite{Salpeter55})
IMF  with M$_{\rm low}$ and M$_{\rm up}$ of 0.1 and 120 M\sunn, respectively.
In Figure \ref{FigModel_E} we plot the calculated evolutionary tracks for
the colours $(U-B), (B-V), (V-R), (R-I)$ and
$(V-I)$. Three types of star formation scenarios are presented:
instantaneous
starburst, constant SFR and exponentially decreasing SFR with the e-fold
time of $\tau$=3~Gyr. The shadowed strips show the $\pm$1~$\sigma$ ranges for
the derived gas-free colours of the underlying stellar population of SBS
0335--052~E at $R_{\rm eff}$ = 7.5\arcsec\ (shown by horizontal lines in the
middles of these strips).
The observed colours of the underlying
component (at $R_{\rm eff}$ = 7.5\arcsec) of the E galaxy
(corrected for the Galactic extinction) and corrected for nebular emission
are presented
in columns 2 and 3, respectively, in the upper half of Table \ref{tabStarAge}.
The respective allowable ages of the stellar
populations for the three SF histories are presented in columns 4, 5 and 6
Table \ref{tabStarAge}.

\begin{table*}
\centering
\caption[]{Ages of the underlying stellar populations in SBS 0335--052 E and W}
\label{tabStarAge}
\begin{tabular}{lrcccc}
\hline \hline
      &              & Nebular      &       \MC{3}{c}{Model stellar ages (Myr)}                         \\
Colour& Observed     & emission     & Instantaneous & Continuous & Exponential       \\
      &              & excluded     &               &            & $\tau =3$~Gyr     \\
\hline
\multicolumn{6}{c}{\bf East galaxy (at $R=$7\farcs5)} \\
$U-B$ & $-0.89\pm0.14$  &  $-0.86\pm0.14$  &  $ 4\div25  $  & $ 4\div120 $  & $ 4\div120$   \\
$B-V$ & $ 0.20\pm0.06$  &  $~~0.10\pm0.07$ &  $100\div350$  & $400\div2000$ & $400\div1500$ \\
$V-R$ & $ 0.19\pm0.07$  &  $~~0.17\pm0.09$ &  $100\div800$  & $200\div6000$ & $200\div3500$ \\
$R-I$ & $-0.30\pm0.11$  &  $~~0.08\pm0.11$ &  $ 14\div120$  & $35\div400$   & $35\div350$   \\
\hline
\multicolumn{6}{c}{\bf West galaxy (at $R=$5\farcs0)} \\
$U-B$ & $-0.39\pm0.16$  & $-0.39\pm0.16$  &  $  45\div200$ & $ >350      $  & $350\div3500 $  \\
$B-V$ & $-0.17\pm0.11$  & $-0.13\pm0.11$  &  $  4\div60 $  & $  4\div250 $  & $  4\div250 $  \\
$V-R$ & $ 0.48\pm0.11$  & $~~0.49\pm0.11$ &  $   >4000   $ & $  >10000$     & $ >10000    $  \\
$R-I$ & $-0.12\pm0.13$  & $-0.03\pm0.13$  &  $  0\div70$   & $0\div160$      & $ 0\div150  $  \\
\hline\hline
\\[-0.2cm]
\multicolumn{6}{l}{Ref: (1) -- Data from this paper;}\\
\end{tabular}
\end{table*}

   \begin{figure*}
   \centering
   \includegraphics[width=5.9cm,bb=33 164 583 709,clip=]{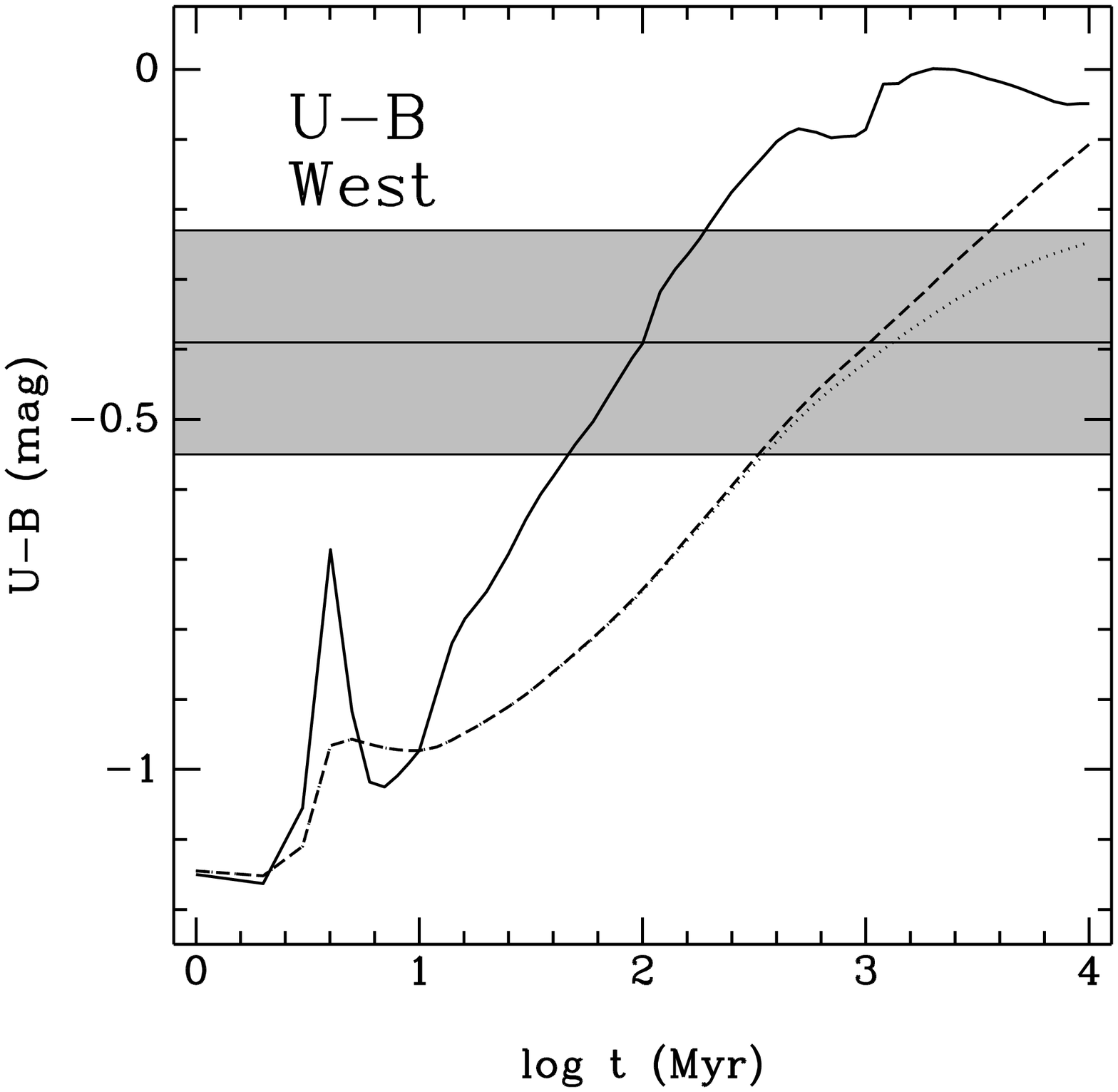}
   \includegraphics[width=5.9cm,bb=33 164 583 709,clip=]{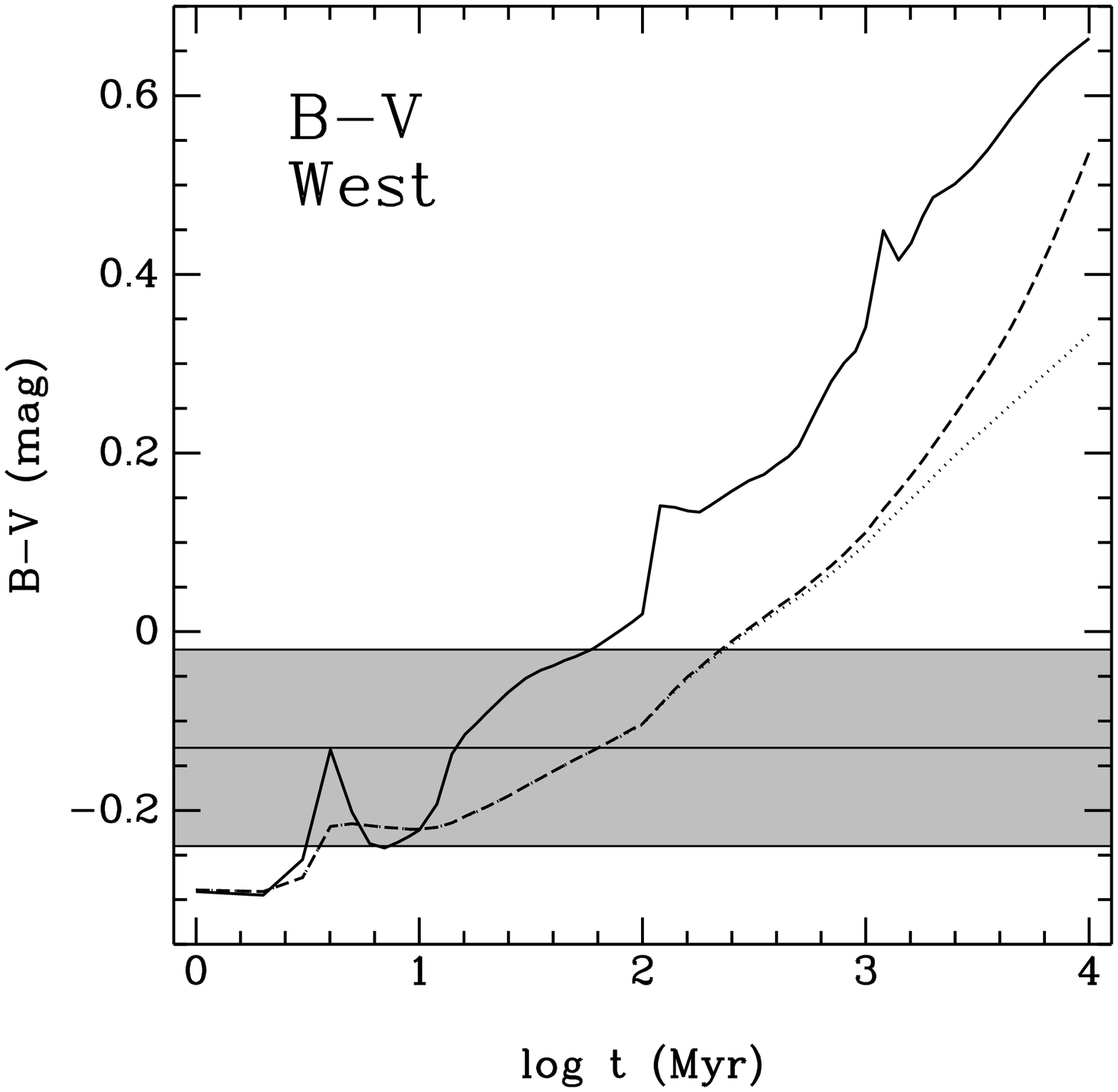}
   \includegraphics[width=5.9cm,bb=33 164 583 709,clip=]{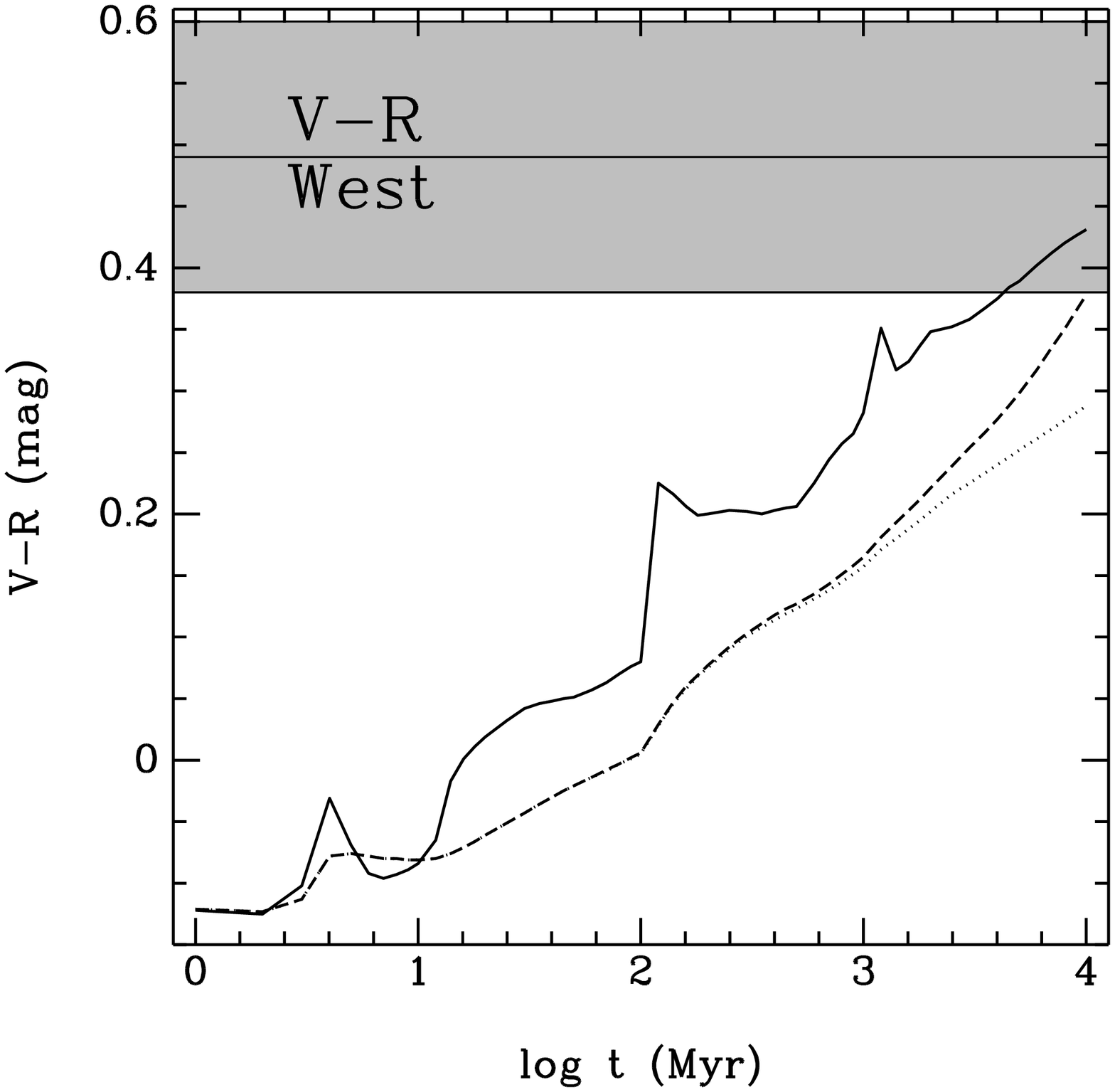}
   \includegraphics[width=5.9cm,bb=33 164 583 709,clip=]{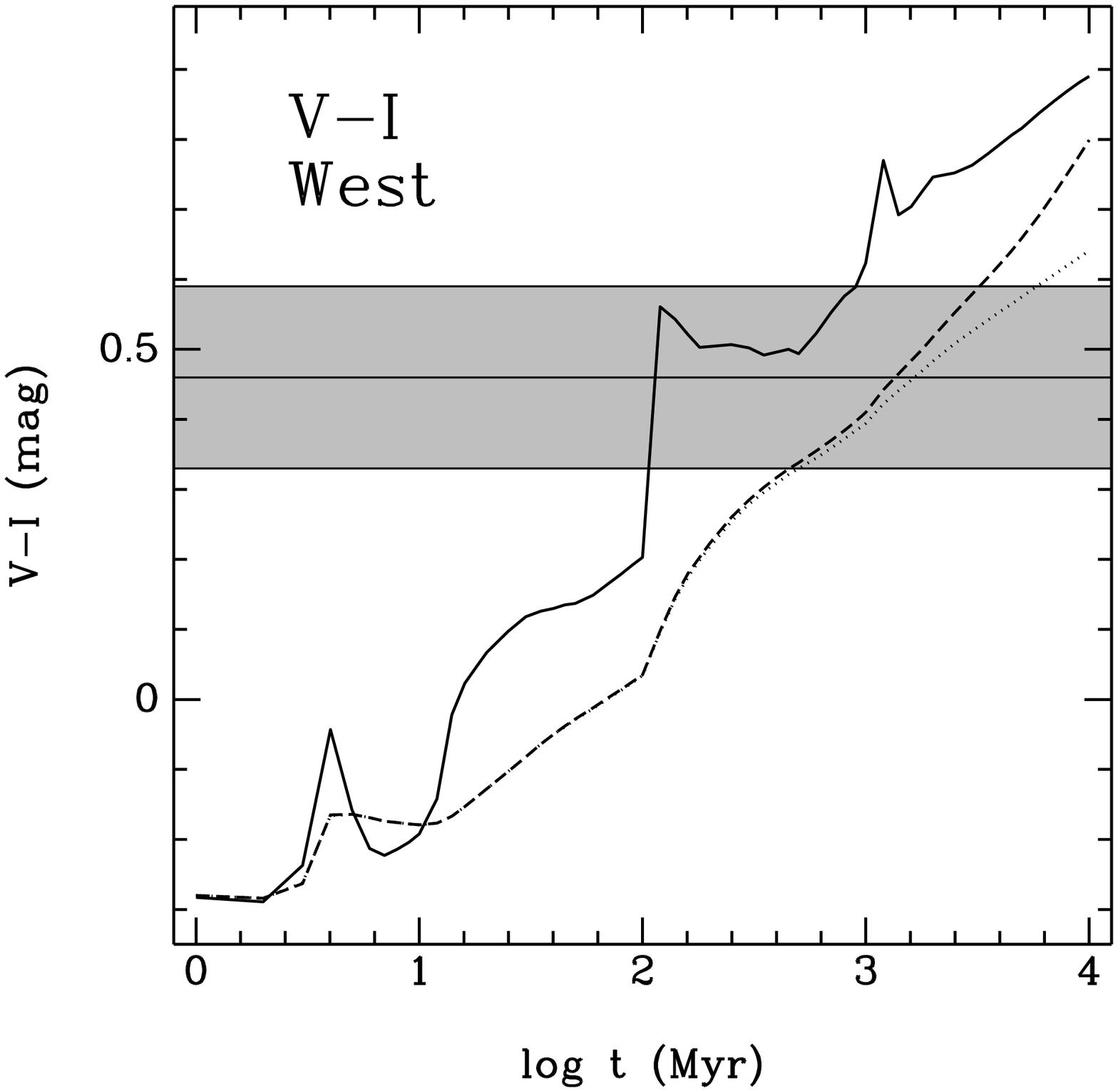}
   \includegraphics[width=5.9cm,bb=33 164 583 709,clip=]{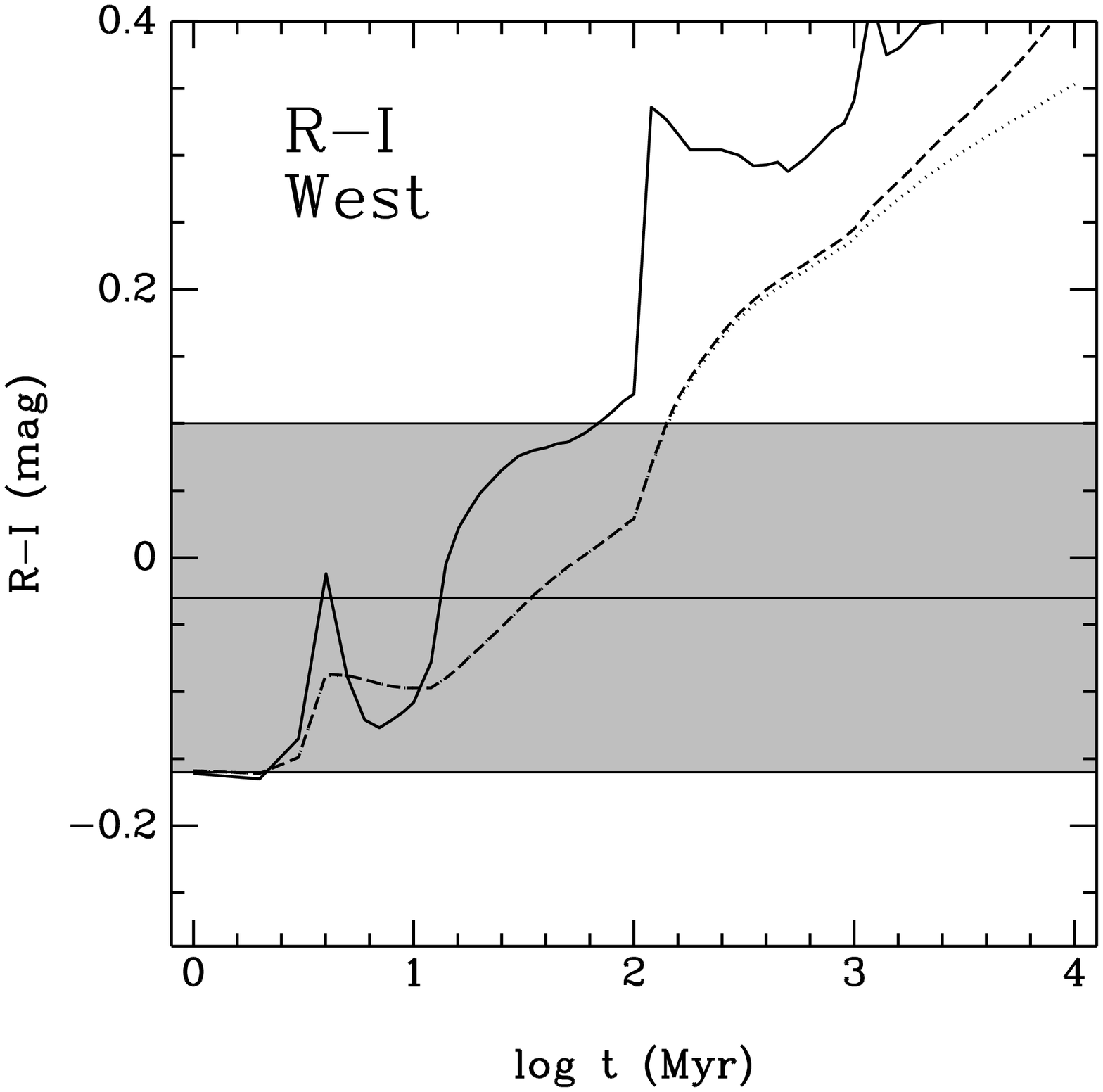}
      \caption{PEGASE.2 model evolution tracks (see text for details).
	 Solid line -- instantaneous starburst model,
	 dotted line -- constant star formation rate,
	 dashed line -- exponentially decreasing
	       SFR with $\tau$=3~Gyr.
	 Horizontal lines show derived colours of the underlying LSB component
	 of the W galaxy with nebular emission subtracted. Shadowed strips
	 show the limits of $\pm$1~$\sigma$ uncertainties of the derived
	 colours.     }
	 \label{FigModel_W}
   \end{figure*}

The similar data on the underlying component (at $R_{\rm eff}$ = 5\arcsec)
of the W galaxy are presented at the bottom of Table \ref{tabStarAge}.
In Figure \ref{FigModel_W}, similar to Figure \ref{FigModel_E}, the gas-free
colours are shown with their uncertainties (shadowed strip), again
compared to the same PEGASE.2 model tracks as for the east galaxy.

\section{Discussion}
\label{Discussion}

\subsection{Comparison of the derived colours with the previous analysis and
 age estimates}

Since the age of the underlying stellar component in
SBS 0335-052~E was the subject of some debate, we discuss this point in more
detail. The $(U-B)$, $(B-R)$ and $(R-I)$ colours of the outermost regions of
this galaxy were presented by Papaderos et al. (\cite{Papa98}). These authors
also compared their results
with the colours of the evolving stellar population, based on
the calculations performed in that work. They concluded that the colours of
the east galaxy, after accounting for the nebular emission contribution, are
consistent with the ages less than $\sim$100 Myr. However, Kunth \&
\"Ostlin (\cite{KO_01}) criticized that conclusion. They
performed a comparison of the colours from Papaderos et al. (\cite{Papa98})
with the more advanced model tracks, based on the PEGASE.2 package. From this
they concluded that the same colours are best explained
by the stellar population with the age of several Gyr.

While for comparison  we also use the models based on the PEGAS.2 package,
our
results differ drastically from those by Kunth \& \"Ostlin (\cite{KO_01}).
The source of this discrepancy is the choice
of colours of the underlying `disk' component.
In Sect. \ref{surface} (Figure \ref{FigColor}) and in Table \ref{tabStarAge}
(second column) we present our estimates of the colours in question.
The latter correspond to those measured for the annulus with $R_{\rm eff}$
= 7\farcs5. Those colours are consistent within the uncertainties with the
colours of the LSB `disk' derived as the respective differences of
$m_{\rm disk}$ for various filters in Table \ref{tabStrParam}
(after correction for the Galactic extinction).
Despite our {\it integrated} magnitudes and colours of this galaxy
being
more or less consistent with those from Papaderos et al. (\cite{Papa98}),
the colours of the `disk' component in the periphery, as given in
Table \ref{tabStrParam}, differ somewhat.
Our colours are significantly bluer. Therefore, even
after the correction for nebular emission, they remain sufficiently blue.
Hence, comparing these gas-free colours of the outermost regions with
the same PEGASE.2 models as used by Kunth \& \"Ostlin (\cite{KO_01}), we
derive the ages of stars much lower compared to those derived by
Kunth \& \"Ostlin
(\cite{KO_01}) from the colours of Papaderos et al. (\cite{Papa98}).

As was noticed above, due to the poor seeing of the $U$-band image, our data
on $(U-B)$ colours are less reliable. The east galaxy `disk'
appears
too blue and is only marginally compatible with the rest colours for
continuous a SF history.
For all other colours of gas-free emission, the best fit is reached for the
instantaneous starburst with an age of $\sim$100 Myr.
For continuous SF laws, different colours give less consistent ages.
However, If such
scenarios are considered as an alternative to the former model,
they give the ages of $\lesssim$400 Myr.
If we accept the $(U-B)$  radial distribution from Papaderos et al.
(\cite{Papa98}), as derived from the data with significantly better seeing,
the compatibility of gas-free $(U-B)$ with other colours improves. Within
the cited uncertainties the ages of the `disk' stars derived from $(U-B)$ will
fall
roughly into the same range of less than a few hundred Myr, as for other
gas-free colours.

Thus, applying the same models to the new colours of the SBS 0335--052~E
`disk'
component, and directly accounting for the observed nebular emission
contribution, we conclude that at the BCG periphery only the young population
gives the detectable contribution to the light of this galaxy.

For the W galaxy there are no self-consistent gas-free colours, which could
match those of a one-mode population (neither instantaneous nor continuous
SF).
All colours, besides $(V-R)$, are rather blue indicating instantaneous SF ages
of $\lesssim$100 Myr,
or a continuous SF with ages less than 200--300 Myr.
$(V-R)$ is, however, too red, implying the minimal age of $\sim$4000 Myr. The
reason for the significant inconsistency between $(V-R)$ and other colours is
unclear.

The only reasonable explanation is related to the possible global error
of the $V$-band zero-point calibration of the order of 0.1 magnitude. We have
checked this option since
the original total $V$-band magnitude of the east galaxy derived from the
BTA image appeared 0\fm2 brighter than that from the HST
measurement (Thuan et al. \cite{TIL97}). We additionally calibrated
our $V$-band image taken with the Jacobus Kapteyn telescope (Canarias). Based on this
calibration, we derived that the total $V$-band magnitude is closer to the
HST value and shifted the zero-point of the BTA image. Despite
the fact that this seems to be the most probable source of the systematic
shift of $V$-band
magnitude in the W galaxy, careful examination of this procedure did not
reveal any sources of the error. Therefore we favor the
colours of the underlying LSB component presented in Table \ref{tabStarAge}.

\subsection{Constraints on the contribution of old stars}
\label{old_stars}

The $M$/$L$  ratio for the old low-mass stars is many times higher than that
of a 100-Myr old `disk' component. Therefore one can not exclude that there
exists a substantial mass of a hidden old stellar population which has
only a subtle contribution to the optical light of the `disk'. To constrain
its mass, we constructed the tracks on the
colour-colour diagrams which describe composite colours of the mixtures of
the 100-Myr `young' SSC with the varying fraction of old stars (similar to
the analysis in Pustilnik et al. \cite{SAO0822}). The relative contribution
of light in the broad-band filters for various mixtures was calculated based
on the PEGASE.2 package. The comparison
of the observed gas-free colours (with their uncertainties) with these tracks
implies that a hidden 10-Gyr old population can have a mass no more than
twice that of the mass of the `young disk'.

In a similar manner, we compared the gas-free `disk' colours with
the model tracks for composite population, in which a `young' component is
defined as that with continuous SF over the last 400 Myr,
while the old one is a result of continuous SF during the last 10 Gyr, as
above. The mass of the hidden old stars should be less  than
one mass of the visible 400-Myr old population. This implies
that if a 10-Gyr old population exists, the SFR related to this should be
a factor of 25 lower than the average SFR during the last 400 Myr.

If we have in mind the hypothesis of merging (see Sect. \ref{merger}),
then the probable 1st close encounter (that occurred about 1 Gyr ago)
might result in the trigger of an earlier SF episode, with an age of
$\sim$0.5 Gyr. The latter case would be consistent with the alternative
interpretation of the `disk' colours as due to  continuous SF during
the last $\sim$400 Myr. The additional argument of the several hundred Myr
age
of the youngest XMD BCGs was suggested recently by Izotov et al.
(\cite{ISGT03}) based on their characteristic value of $\log$(N/O) = --1.6
compared to the lowest value observed in the damped Ly-$\alpha$ (DLA) systems  of
$\log$(N/O) = --2.3.

\subsection{Ionized hydrogen morphology}

Quite poor seeing prevents us from resolving the compact starburst
component of the east galaxy.
However,  our rather deep H$\alpha$-image allows us to follow the
morphology of
the ionized gas in the outermost low contrast regions. This is
very irregular and filamentary. A very prominent arc is seen at $\sim$5\arcsec\
NW the H$\alpha$ brightness peak. At larger distances some filaments
also emerge along this direction.

There is a
small displacement of the maximum of H$\alpha$ brightness (and EW(H$\alpha$))
for the east galaxy relative to the $B$-band brightness peak
(Fig. \ref{FigColor}).
On the other hand, the
brightness peaks of the east galaxy in all broad-band filters
coincide within $\sim$0\farcs14. The same is valid for the W galaxy.
The H$\alpha$ peak in the east galaxy is displaced from that in $B$-band by
0\farcs34 to the East and 0\farcs56 to the South, a total
of 0\farcs65 (or 167 pc).
A similar displacement of the peaks in the
distributions of continuum and strong lines in the long-slit
spectra was noticed by Izotov et al. (\cite{ILC97,Izotov99}).
Most probably this is related to the complex structure of the region of
current/recent star formation, as seen from the HST $V$-band image
of Thuan et al. (\cite{TIL97}). The total extent was
$\sim$2\arcsec\ along the direction SSE--NNW; it was resolved into six
super-star clusters (SSCs) with ages from a few Myr to 25 Myr.

For the W galaxy, the position of the H$\alpha$ peak coincides with that of
the $B$-band within 0\farcs17, that is, the morphology of the ionized gas
correlates rather well with that seen in the broad-band filters.
This presumably indicates significantly lower age
differences between the star-forming regions in the two components of the
W galaxy.

\subsection{SFR, stellar and gas masses}
\label{SFR}

Having the H$\alpha$-luminosity of the studied galaxies, we could derive its
SFR, following commonly used relations. But this is not of practical use.
As shown by Weilbacher \& Fritze-v.-Alvensleben (\cite{WF01}, hereafter WF01),
the
relation between H$\alpha$-luminosity and SFR in the blue compact galaxies is
strongly time-dependent (on the time scale of a few Myr).
This differs from the relations
of Hunter \& Gallagher (\cite{HG86}) or Kennicutt et al. (\cite{Kennicutt94})
derived for dIrr or spiral galaxies with more or less stationary star
formation and metallicity near solar. For short intense starbursts
taking place in low-metallicity BCGs, the SFR derived from H$\alpha$
luminosity through the commonly used formula can result in an overestimation
by a factor of $>>$ 10 (WF01).

More important parameters are the total mass of stars formed
during the current SF episode and the time elapsed since the beginning of this
event. When the latter is known, the total mass of a starburst can be
found using its total luminosity in broad bands and Balmer lines, through,
e.g., the Starburst99 (SB99, Leitherer et al. \cite{SB99}) or PEGASE.2
packages.
Adopting the EWs of H$\alpha$ and H$\beta$ from Izotov et al.
(\cite{ILC97,Izotov99}) and Lipovetsky et al. (\cite{Lipovetsky99}) for the
east and west galaxies, respectively, we can estimate the age of instantaneous
starbursts in these objects.

For the east galaxy, EW(H$\beta$)$\sim$240~\AA\ near the central region implies
the age of a starburst with the Salpeter IMF (in the range
between $M_{\rm low}$=0.8 and $M_{\rm up}$=120 $M$\sunn) of either $\sim$3 Myr
or 3.7 Myr (Schaerer \& Vacca \cite{SV98}).
Discovery of WR features in the brightest part of its spectrum (Izotov et al.
\cite{Izotov01}) corroborates this estimate, since the models predict the
appearance of WR stars at such low metallicities only in the range of ages
between $\sim$3 and $\sim$4 Myr (Schaerer \& Vacca \cite{SV98}).
Then, for $m_{\rm burst}$=17.49 (from column 8 of Table
\ref{tabStrParam}) we get its absolute magnitude $M^{B,0}_{\rm burst}=-$16.38.
Respectively, $M^{B,0}_{\rm disk}=-$15.91. From the PEGASE.2 models with
the Salpeter IMF and $M_{\rm low}$ and $M_{\rm up}$ of 0.1 and 120 $M$\sunn,
for the adopted starburst $B$-band luminosity and $Z$=$Z$\sunn/50, the mass
of the current starburst $\sim$5.7$\times 10^{6}~M$\sunn. For the mass of the
underlying disk, if it was formed in the instantaneous starburst $\sim$100
Myr ago, $M_{\rm disk}$$\sim$4.2$\times 10^{7}~M$\sunn. For the alternative
option of the underlying disk formed continuously over the last 400 Myr,
with the same IMF and $Z$, its mass would be 4.0$\times 10^{7}~M$\sunn.

For the west galaxy the  EW(H$\beta$)$\sim$100 and 140
\AA\ measured in the 2 knots (Lipovetsky et al. \cite{Lipovetsky99})
imply the larger age of recent starburst, $\sim$4 Myr. However, as
follows from the comparison of the H$\alpha$ flux measured here with that from
Lipovetsky et al. (\cite{Lipovetsky99}), about a half of all H$\alpha$ flux
is outside the region for which the EW(H$\beta$) was derived. This implies
that the real EW(H$\beta$), which should be compared with the models, is about
a factor of 2 higher. This, in turn, implies that the age of the recent SF
episode in the W galaxy is also lower, $\sim$3.8 Myr. With the use of the same
PEGASE.2 models, the masses of the recent starburst and the underlying older
population are derived  from the absolute blue magnitudes for the starburst
and for the `disk'
in Table \ref{tabStrParam}.
They are $\sim$1.3$\times 10^{5}~M$\sunn\  for the starburst and
$\sim$1.2$\times 10^{7}~M$\sunn\ for the
`disk' with the characteristic age of $\sim$100 Myr.
The total visible masses of stars in the east and west galaxies differ by a
factor of $\sim$4.
For the option of the `disk' with continuous SF during the last $\sim$400 Myr
the underlying disk mass of the W galaxy would be of
$\sim$1.25$\times 10^{7}~M$\sunn.

The gas or stellar mass-fractions are important parameters in models of
chemical evolution. We address this issue below.
If we consider the \ion{H}{i} mass for each of the galaxies separately,
then the mass of neutral gas related to each of them is $\sim$1.0 and 1.1
$\times$10$^{9}~M$\sunn\ (Pustilnik et al. \cite{VLAdata}).

The mass of the ionized gas in the east and west galaxies can be roughly
estimated from their total H$\alpha$ luminosities, assuming the gas spatial
distribution.
For `disk' fractions of both galaxies we assumed the scaleheight to be
close to their scalelengths. For the `Gaussian' component in the east galaxy
we
assumed a radial distribution with the characteristic radius from Table
\ref{tabStrParam}. With these inputs we estimate the ionized gas mass as
2.5$\times$10$^{8}~M$\sunn\ for the east galaxy and
2.4$\times$10$^{7}~M$\sunn\ for the west galaxy.

The mass fraction of visible stars in the east galaxy  comprises
$\sim$0.035 of the total baryon mass, while in the west galaxy this is
only $\sim$0.011, implying rather different SF histories.
For the option of the underlying `disks' with the age of $\sim$400 Myr,
the mass fractions of stars will be close to the above values.
We emphasize
also other important differences in the properties of these two galaxies.
As noticed in Sect. \ref{surface}, while the ages of current starbursts
are very close, their relative strengths differ by a factor of 8.
The significant difference in their central surface brightnesses
indicates an intrinsic difference in their mass distribution and/or
SF history.

\subsection{SBS 0335--052 system as a representative
	       of the merger evolution sequence}
\label{merger}

As was discussed, e.g., by Pustilnik et al. (\cite{VLAdata}),
there are two  scenarios resulting in such an unusual pair
of dwarf galaxies. The first one is related to a unique retarded protogalaxy,
a kind of free-flying \ion{H}{i} cloud, which  could
appear as a local Ly-$\alpha$ absorbing cloud (e.g., Manning
\cite{Manning02,Manning03}).
This large neutral gas cloud finally reached outskirts of the loose group
LGG 103 and experienced sufficiently strong tidal disturbance from the giant
spiral NGC 1376 (at a projected distance of $\sim$150 kpc).
Resulting gas collapse and star formation occurred recently in two off-center
positions roughly symmetric relative to the cloud center. They appear now as
two young galaxies immersed in the single disturbed \ion{H}{i} cloud.

The second scenario considers two very gas-rich objects
with similar gas mass
(either purely gas clouds, or very slow evolving galaxies, as expected for
very low SB galaxies) currently seen in \ion{H}{i} morphology to be in
contact. Their recent collision and the consequent merging caused strong
disturbance and the loss of gas stability in each of them (see, e.g.,
simulations in Springel (\cite{Springel00}) and references therein).
Gas infall to the
component centers with subsequent collapse caused the onset of the
first or the additional enhanced star formation.

The
very significant difference in the optical properties of the east and west
galaxies, discussed above,
suggests that these two objects are intrinsically very different and, hence,
favors the second scenario. We discuss this option in more
detail and follow possible interrelations of SBS 0335--052 with other XMD
galaxies.

One of the main objections for the collision/merging scenario is very
small probability of such an event, given the estimates of the
density of such objects. The argument which completely diminishes this
difficulty is the existence of another similar system. This is well known
Dw 1225+0152, the optical counterpart of the giant \ion{H}{i} cloud
HI 1225+01 with $Z \sim Z$\sunn/20 studied in detail by Salzer et al.
(\cite{Salzer91}) and  Chengalur et al. (\cite{Chengalur95}).
This \ion{H}{i} cloud  consists of two separate components  with
about the same
mass ($\sim$10$^{9}~M$\sunn) at a distance of $\sim$2 visible diameters
of these two \ion{H}{i} components, joined by tidal bridge.
While in the NE component we observe a star-bursting dwarf galaxy,
the SW component
shows lack of detectable optical emission down to a $V$-band surface
brightness of 27.0 mag~arcsec$^{-2}$ (Salzer et al. \cite{Salzer91}).
The SW component is thus either a purely gas object, or an extremely low
surface brightness galaxy. The stability of gas in the SW component is
significantly higher than for the NE one, since the current SF rates in both
objects differ drastically, while their tidal disturbances
are comparable.

Comparing the \ion{H}{i} morphology and
velocity field of this object with the N-body simulations (Barnes \&
Hernquist \cite{Barnes91}), Chengalur et al. (\cite{Chengalur95}) argue
that these objects have recently experienced the first close encounter, and
during the next one the system will merge. Consequently, while the merging
of the two components will progress, one can expect that the stronger
disturbance will cause the gas collapse in the SW component and trigger its
star formation. This seems to resemble  the current situation
in the SBS 0335--052 system, where we see strong starburst in the east
component, and quite a modest SF event in the west component.
One more similarity of these
two systems is their belonging to low-density outskirts of galaxy
aggregates. The distance to Dw 1225+0152 is somewhat uncertain due
to the complications with the distance estimates in the vicinity of the
Virgo cluster.  However, it is quite probable that it is situated at
the distant periphery of this cluster.

If this is the case, the existence of both systems implies that in the
near environment of various bound galaxy aggregates (or more exactly, in
some fraction of them) there exists an abundant population of very gas-rich
objects with masses of the order of 10$^{9}$ $M$\sunn. Presumably, such
galaxy aggregates are quite rare, since searches for \ion{H}{i}
clouds in the Local Volume and some nearby groups with the mass detection
limit of 10$^{7}$--10$^{8}~M$\sunn\ gave only a few detections (e.g.,
Henning \&
Kerr \cite{Henning89}, Putman et al. \cite{Putman98}).

If this scheme is correct, then SBS 0335--052 and Dw
1225+0152 can be considered as an evolution sequence along with two other
recently discovered XMD blue compact galaxies: HS 0822+3542 ($Z \sim$
$Z$\sunn/30, Kniazev et al. \cite{Kniazev00b}) and HS 0837+4717 ($Z \sim$
$Z$\sunn/20, Kniazev et al. \cite{Kniazev00a}).
HS 0822+3542 is a member of a binary gas-rich galaxy system with a very small
relative velocity and projection distance of 11 kpc, or $\sim$4 components'
\ion{H}{i} diameters (Pustilnik et al. \cite{SAO0822}, Chengalur et al.
\cite{GMRT}).  The mass and luminosity of the second component are comparable
to those of the BCG, but its surface
brightness and SF are typical of LSB dwarfs. This system can be considered
as a pre-merger, preceding the stage observed in HI 1225+01 system.

HS 0837+4717 seems a very probable well advanced
merger,
with a disturbed morphology and gas velocity field. Two knots near the
object center are separated by only $\sim$2 kpc (Pustilnik et al.
\cite{HS0837}). In the suggested scheme, the
case of HS 0837+4717 can be treated as a well advanced system of SBS
0335--052. Two more XMD galaxies -- HS 0122+0743 and HS 2236+1344, found
in HSS-LM (Ugryumov et al. \cite{HSS_LM}) also support
the merger scenario.
It is too early to estimate the fraction of mergers among XMD BCGs.
However, it is worth noting that the \ion{H}{i} morphology and kinematics
of a prototypic XMD galaxy I~Zw18 (van Zee et al. \cite{vanZee98})
suggest a possible recent interaction/merger in this system. This scenario
can be examined more carefully by comparing all available observational
data with N-body simulations of low-mass gas-rich (proto)galaxy mergers.

Such an evolutionary scheme does not imply the tight correlation of
metallicity
with the merger phase, since it can invoke a rather wide range of very low
metallicities of pre-merger objects.
However, on average, the objects representing
almost complete mergers should be more metal-enriched due to the intense SF
accompanying the merger at the final stage.

Within the merger scenario some additional implications can be made for
SBS 0335--052 system itself. If this is a well-advanced merger, presumably
at the stage of the second (and the last) encounter, some interesting
consequences should exist of the first encounter, similar to
that seen in the system HI 1225+01. Following the analogy between these two
systems, we
could expect that during the first encounter the significant tidal disturbance
ignited some SF in the E component, which is characterized by a higher
surface brightness of the underlying `disk'. In Sect. \ref{old_stars}
we estimate the mass of stars formed in this SF episode.

One problem with such a scenario is the observed close metallicity of
gas in the two merging dwarfs. This likely can be explained by the
significant exchange of
gas between the two components during the first encounter (e.g., Mihos \&
Hernquist \cite{MH96}) and the following mixing, which should result in
similar metallicities.
The best way to better constrain when the first episode of star formation
took place and how much gas mass was transformed into stars
is to study the colour-magnitude diagram of individual resolved
stars on the galaxy periphery.
In such a scenario, if the east galaxy is truly young, the west galaxy also
should be young, since this is even more stable relative to the external
perturbations.

\section{Summary and conclusions}
\label{conc}

In this work we have analyzed the results of the 6-m telescope $UBVRI$ and
H$\alpha$ surface photometry of two extremely
metal-deficient gas-rich galaxies SBS~0335--052~E and W.
They represent a unique galaxy pair, both components of which
show the properties of candidate young galaxies. We subtract
the nebular emission, using their H$\alpha$-line intensity maps and
the additional data on the spectra of the ionized gas. This gives us the
gas-free colours of the underlying stellar population outside the regions
of the current star-formation bursts. These colours are significantly
bluer than those claimed in earlier works.
These gas-free colours are compared with various PEGASE.2 evolution model
tracks. The most self-consistent estimates of the ages of underlying stars
do not contradict the hypothesis of youth of both galaxies.

The evidence for the importance of galaxy interactions and mergers, in
particular, for the star-forming activity in BCGs was presented, e.g., by
Pustilnik et al. (\cite{PKLU01}). The arguments for the merger nature of
luminous
BCGs were presented by Bergvall \& \"Ostlin (\cite{BO02}, and references
therein). In this aspect, XMD galaxies with SF bursts show similarity to the
majority of more typical BCGs. A significant fraction of SF bursts in these
galaxies are triggered due to strong disturbance from the other galaxy (or
protogalaxy).

We draw the following conclusions:
\begin{enumerate}
  \item The total magnitudes of new deep $UBVRI$  photometry of SBS
    0335--052~E are consistent with the earlier data from Papaderos
    et al. (\cite{Papa98}). This gives additional confidence in the new
    data for the Western  galaxy derived from the same CCD frames.
    The only possible exception is its suspicious $V$-band zero-point.
  \item The colours of the outer regions for the Eastern galaxy, derived in
	this work, however, are significantly bluer than those claimed in
	earlier works. Similarly, our $(U-B)$ and $(R-I)$ colours of the
	outer regions of the Western galaxy are also significantly bluer than
	those from earlier studies.
   \item The gas-free colours of the outer regions in the Eastern galaxy
    are derived with the use of the H$\alpha$-intensity maps.
    Compared to the evolution models of stellar populations from PEGASE.2,
    they are consistent (exccept for the more uncertain $(U-B)$ colour) with
    the     ages of
    $\lesssim$100 Myr for instantaneous starburst. For a less consistent
    but still acceptable option -- a scenario with continuous SF, the
    duration of this SF episode
    would be $\lesssim$400 Myr. For the underlying disk with an age
    of $\lesssim$100 Myr, the mass of a 10-Gyr old
    stellar population  cannot exceed twice that inferred for the
    young disk stellar population.
    If, however, the disk was forming during the last $\lesssim$400 Myr,
    the mass of the hidden 10-Gyr old population would be less than or
    equal to that of the young stellar disk.
   \item For the gas-free colours of the outer regions of the W galaxy,
     according to the data obtained the conclusions are less
     certain. The fully self-consistent combination of colours, matching the
     model colours of a stellar population with a unique SF history, is
     absent. However, 3 of the 4 independent colours are consistent with
     an instantaneous SF burst with an age of $\sim$70 Myr. Similarly,
     for the less consistent case of continuous SF, 3 of 4 colours indicate
     a period of star formation less than 200--300 Myr.
   \item Star-forming properties and the central surface brightnesses of
    the E and W
    galaxies differ drastically, suggesting that the two galaxies are
    intrinsically very different. This, in turn, favors the hypothesis
    of the collision of two gas-rich XMD galaxies/protogalaxies.
   \item We suggest an evolutionary scheme in which the systems
    HS 0822+3542/SAO 0822+3545, HI 1225+01, SBS 0335--052 and HS 0837+4717
    comprise the sequence of XMD gas-rich objects at various stages of
    merging (from pre-merger to full merger). The interaction-induced
    SF in XMD gas-rich galaxies seems to be an important channel of their SF
    bursts.
\end{enumerate}

\begin{table}
\caption[]{Main parameters of galaxies SBS~0335--052~E and W}
\label{tabMain}
\begin{tabular}{lrrc} \hline\hline
Parameter                              & 0335$-$052~E        & 0335$-$052~W     & Ref   \\
\hline
\\[-0.3cm]
$\alpha_{\rm 2000}$                 & $03^{\rm h} 37^{\rm m} 44\fs 03$ & $03^{\rm h} 37^{\rm m} 38\fs 40$ & \\
$\delta_{\rm 2000}$                 & $-05\degr 02\arcmin 38\farcs8$         & $-05\degr 02\arcmin 36\farcs 4$        & \\
$B_{\rm tot}$ [mag]                 & $16.95\pm0.03$        & $19.14\pm0.03$       & 5   \\
$V_{\rm hel}$ [\kms]                   & 4057                &  4017               & 2   \\
$D$ [Mpc]                             & 54.3                    & 54.3             & 1   \\
$A_{\rm B}$ (NED)                      & 0.20                  & 0.20              &      \\
$M_{\rm B}^{0}$ [mag]                  & $-16.92$               & $-14.73$         & 5    \\
Size             [arcsec]              & $23\times20$          & $14\times14$      & 3   \\
12+$\log$(O/H)    [dex]                 &  7.29                 &  7.22             & 4   \\
$T_{\rm e}$(\ion{O}{iii}) [K]           & 20290               & 17200               & 4   \\
$M($\ion{H}{i}$) [\times10^{9}M_{\sun}$]              & 0.80   & 0.89              & 2   \\
$M($\ion{H}{i}$)/L_{\rm B}$ [$M$\sunn/$L$\sunn]    & 0.9    & 7.3               & 5   \\
\\[-0.3cm]
\hline\hline
\\[-0.3cm]
\multicolumn{4}{l}{(1) -- Papaderos et al. (\cite{Papa98});
		   (2) -- Pustilnik et al. (\cite{VLAdata});} \\
\multicolumn{4}{l}{(3) -- maximal extent on $B$ isophote 27.5 mag\,arcsec$^{-2}$ from} \\
\multicolumn{4}{l}{Papaderos et al. (\cite{Papa98}) and this work;
		   (4) -- Izotov et al. } \\
\multicolumn{4}{l}{(\cite{Izotov99}) for the Eastern and Lipovetsky et al. (\cite{Lipovetsky99}) for the} \\
\multicolumn{4}{l}{Western galaxy; (5) -- this work.}
\end{tabular}
\end{table}

\begin{acknowledgements}
The authors thank A.Kopylov for the help in observations. We are grateful to
Y.Izotov who provided the deep long-slit Keck telescope  spectrum of SBS
0335--052E, used in this work as a reference to model ionized gas emission.
We are thankful to Y.Izotov for the fruitful discussion and useful notes on
the preliminary version of the paper. The authors acknowledge the useful
suggestions made by the anonymous referee, that allowed them to improve the
presentation.
The work was partly supported by the Russian Federal program "Astronomy".
This research has made use of
the NASA/IPAC Extragalactic Database (NED) which is operated by the Jet
Propulsion Laboratory, California Institute of Technology, under
contract with the National Aeronautics and Space Administration.
\end{acknowledgements}

\clearpage
\end{document}